\documentclass[apj]{emulateapj}

\usepackage{graphicx}
\usepackage{natbib,times}
\usepackage{verbatim}

\newcommand{\be}{\begin{equation}}
\newcommand{\ee}{\end{equation}}

\newcommand{\msun}{{$M_{\odot}$}}

\newcommand{\gtsima}{$\; \buildrel > \over \sim \;$}
\newcommand{\ltsima}{$\; \buildrel < \over \sim \;$}
\newcommand{\prosima}{$\; \buildrel \propto \over \sim \;$}
\newcommand{\gsim}{\lower.5ex\hbox{\gtsima}}
\newcommand{\lsim}{\lower.5ex\hbox{\ltsima}}
\newcommand{\simgt}{\lower.5ex\hbox{\gtsima}}
\newcommand{\simlt}{\lower.5ex\hbox{\ltsima}}
\newcommand{\simpr}{\lower.5ex\hbox{\prosima}}

\newcommand{\etal}{{et al.~}}
\newcommand{\cxo}{{Chandra}}

\newcommand{\logedd}{log($L_{\rm X}$/$L_{\rm Edd}$)}

\newcommand{\mstar}{$M_{\star}$}

\begin{document}

\title{AMUSE-Field II. Nucleation of early-type galaxies in the field vs. cluster environment}

\author{Vivienne F. Baldassare\altaffilmark{1}, Elena Gallo\altaffilmark{1}, Brendan P. Miller\altaffilmark{1,2}, Richard M. Plotkin\altaffilmark{1}, Tommaso Treu\altaffilmark{3}, Monica Valluri\altaffilmark{1}, Jong-Hak Woo\altaffilmark{4}}
\altaffiltext{1}{Department of Astronomy, University of Michigan, Ann Arbor, MI 48109}
\altaffiltext{2}{Physics and Astronomy Department, Macalester College, Saint Paul, MN 55105}
\altaffiltext{3}{Physics Department, University of California, Santa Barbara, CA 93106}
\altaffiltext{4}{Astronomy Program, Department of Physics and Astronomy, Seoul National University, Seoul, Republic of Korea}

%\date{present}							% Activate to display a given date or no date
\begin{abstract}

The optical light profiles of nearby early type galaxies are known to exhibit a smooth transition from nuclear light deficits to nuclear light excesses with decreasing galaxy mass, with as much as 80 per cent of the galaxies with stellar masses below $10^{10}$ \msun\ hosting a massive nuclear star cluster.  
At the same time, while all massive galaxies are thought to harbor nuclear super-massive black holes (SMBHs), observational evidence for SMBHs is slim at the low end of the mass function. 
Here, we explore the environmental dependence of the nucleation fraction by comparing two homogeneous samples of nearby field vs. cluster early type galaxies with uniform \textit{Hubble Space Telescope} (HST) coverage. Existing \textit{Chandra X-ray Telescope} data for both samples yield complementary information on low-level accretion onto nuclear SMBHs. 
Specifically, we report on dual-band (F475W \& F850LP) Advanced Camera for Surveys (ACS) imaging data for 28 out of the 103 field early type galaxies that compose the AMUSE-Field Chandra survey, and compare our results against the companion HST and Chandra surveys for a sample of 100 Virgo cluster early types (ACS Virgo Cluster and AMUSE-Virgo surveys, respectively). 
We model the two-dimensional light profiles of the field targets to identify and characterize NSCs, and find a field nucleation fraction of $26\%^{+17\%}_{-11\%}$ (at the 1$\sigma$ level), consistent with the measured Virgo nucleation fraction across a comparable mass distribution ($30\%^{+17\%}_{-12\%}$). 
Coupled with the Chandra result that SMBH activity is higher for the field, our findings indicate that, since the last epoch of star formation, the funneling of gas to the nuclear regions has been inhibited more effectively for Virgo galaxies, arguably via ram pressure stripping.
\end{abstract}
\section{Introduction}

The assembly and merging history of supermassive black holes (SMBHs) at the centers of massive galaxies appears to proceed in close connection with -- and possibly even regulate -- the growth of their host galactic bulges.   Perhaps the most well known incarnation of this is the ``$M_{\rm BH}$-$\sigma$'' relation, or the correlation between the mass $M_{\rm BH}$ of a SMBH and the velocity dispersion $\sigma$ of the bulge (\citealt{2000ApJ...539L...9F}, \citealt{2000ApJ...539L..13G}, \citealt{Tremaine:2002dq}, \citealt{2009ApJ...698..198G}, \citealt{2013ApJ...764..184M}; see also \citealt{2010ApJ...716..269W,2013ApJ...772...49W} for M-$\sigma$ relation for active galactic nuclei).  Tight scaling relations have also been claimed between SMBH mass and bulge mass/luminosity (\citealt{Marconi:2003fk}, \citealt{Haring:2004lr}).
While the number of reliable dynamical measurements for black hole mass has grown by a factor of about 5 over the last decade or so, it remains unclear whether these power law scaling relations break down at the highest and lowest masses (\citealt{Lauer:2007ys}, \citealt{Greene:2010fr}, \citealt{Kormendy:2013ve}), and whether classical vs. pseudo-bulges lead to different scaling relations (see e.g. \citealt{Jiang:2011vn}, \citealt{Kormendy:2011rr}).  Related to the above issues, the black hole mass function itself is largely unconstrained at low masses (\citealt{Greene:2007kx}, \citealt{Kelly:2010qy}), and its determination may in turn be biased by the assumption of a single power-law relation with log-normal scatter \citep{Kelly:2012rt}.  

Based on detailed morphological analysis of a large sample of nearby early type galaxies spanning over four decades in stellar mass ($M_{\star}$), \cite{:uw} proposed that, for galaxies with $M_{\star}$ less than a few times $10^{10} M_{\odot}$, compact stellar nuclei -- with half-light radii between 2-5 pc and about 20 times brighter than typical globular clusters \citep{2004AJ....127..105B} -- may take over from SMBHs as the dominant form of mass aggregation in galactic nuclei (a similar result has been reported for a comparably large sample of spiral galaxies by \citealt{Rossa:2006pd}). 

Most relevant to this paper, detailed work on the occurrence rate and properties of these ``nuclear star clusters" (NSCs; \citealt{2004AJ....127..105B}; \citealt{2005ApJ...618..237W}; \citealt{Carollo:1998uq}; \citealt{Matthews:1999kx}; \citealt{Balcells:2003vn}) was carried out within the Advanced Camera for Surveys Virgo Cluster Survey (ACS VCS; \citealt{2004ApJS..153..223C}), consisting of dual-band {\it Hubble Space Telescope} (HST) ACS observations of 100 early-type galaxies in the Virgo Cluster.  Later augmented by the Fornax Cluster Survey \citep{2007ApJS..169..213J}, the ACS VCS showed that between 66\% and 82\% of early-type galaxies with absolute B-magnitude $M_{\rm B} < -15 $  host nuclear star clusters, a much larger percentage than had been thought based on ground observations (\citealt{2006ApJS..165...57C}, \citealt{:fz}).   \cite{2007ApJ...671.1456C} confirmed that NSCs preferentially reside in galaxies with $-19.5 < M_{\rm B} < -15 $ mag, and found that there is a smooth transition from nuclear light deficit to nuclear light excess with decreasing galaxy luminosity.  
Also using the ACS VCS data, \cite{:uw} found that the masses of NSCs correlate with the virial masses of their host spheroidal galaxies, and that this relation extends from the scaling relation between SMBH masses and the bulge masses of their host galaxies, possibly indicating a common growth mechanism for NSCs and SMBHs, and perhaps a shared formation mechanism.  A common scaling relation for NSCs and SMBHs was also found independently by \cite{Wehner:2006cr}.  These results would be consistent with a scenario whereby SMBHs are the dominant -- perhaps sole -- mode of nuclear mass aggregation at the center of bright massive galaxies, becoming progressively less common down the mass function and disappearing entirely at the faint end, to be replaced by NSCs (the existence of a common scaling relation for the SMBHs and the NSCs has been by challenged by \citealt{2012MNRAS.422.1586G} and \citealt{2012MNRAS.424.2130L}). 

Over the last few years, an increasing number of galaxies hosting both NSCs and SMBHs have been identified \citep{2009MNRAS.397.2148G, Neumayer:2012yq, :um}.
In particular, \cite{2008ApJ...678..116S} took a somewhat different approach from previous studies by searching for active galactic nuclei (AGN) in 176 galaxies that were previously known to contain NSCs.  Based on their analysis, at least 10\%  of their sample -- spanning a wide range in masses and Hubble types -- hosts both NSCs and SMBHs, strongly suggesting that galaxies harboring NSCs ``have AGN fractions consistent with the population of galaxies as a whole''. 
In order to better understand the connection between these objects, as well as their respective formation mechanisms, it is desirable to undertake systematic studies that characterize both NSCs and SMBH activity over a sample that is unbiased with respect to nuclear properties.  

With the goal of delivering the first unbiased census of low-level SMBH activity in the local universe, the AMUSE-Virgo (AGN Multi-wavelength Survey of Early-Type Galaxies;  \citealt{:kj}, \citealt{:um}, \citealt{:fg}, \citealt{Leipski:2012rt} ) survey acquired \textit{Chandra X-ray Telescope} ACIS (Advanced CCD Imaging Spectrometer) observations for all 100 early-type galaxies targeted by the ACS VCS.  As the survey was tailored to probe down to Eddington scaled X-ray luminosities as low as \logedd $\simeq -9$, it provides a relatively inexpensive (i.e., compared to dynamical studies) way to identify SMBHs in formally ``inactive" nearby galactic nuclei.  While, in the absence of a NSC, the issue of potential contamination to the nuclear X-ray signal from bright low mass X-ray binaries\footnote{The sample is comprised of early type galaxies only, ensuring negligible contamination from high mass X-ray binaries. } (LMXBs) can be addressed quantitatively based on the known shape and normalization of the X-ray luminosity function of LMXBs (Gilfanov 2004), the presence of a NSC demands a more conservative treatment (see Gallo \etal 2010 for a detailed discussion).  
For the VCS sample, the combined \cxo\ and HST data indicate that between 24\% and 34\% of the targeted galaxies host a {\it bona fide} X-ray active SMBH. The fraction of hybrid nuclei, hosting both a SMBH and a NSC is estimated between 0.3\% and 7\% for \mstar\ below $10^{11}$\msun\ and to be lower than 32\% above it (at the 95\% confidence level; \citealt{:um}).

Born as an extension of AMUSE-Virgo, the AMUSE-Field \cxo\ survey (\citealt{:zr}, \citealt{:fg}, \citealt{Plotkin:2014vn} on ULXs, Miller \etal 2014, submitted) was designed to deliver the first measurement of low-level SMBH activity in a field environment.  AMUSE-Field targeted a volume-limited sample of 103 nearby field early types spanning over three orders of magnitude in host stellar mass;  analogous to the Virgo sample, the field galaxies were selected based solely on optical properties as classified by HyperLeda \citep{2003A&A...412...45P}, to create a sample that is unbiased with respect to nuclear properties.  For the field sample, 45\% $\pm$ 7\% of targets were found to host an X-ray active SMBH \citep{:zr}.  However, this measurement relies on the assumption that the fraction of field objects hosting a NSC is the same as that found for the AMUSE-Virgo targets.  In order to properly compare the incidence of SMBH activity as well as stellar nucleation as a function of host stellar mass for the field sample as well as for Virgo, we acquired dual-band HST/ACS observations for a subsample of the AMUSE-Field galaxies.   

Combined, the Chandra/ACIS and HST/ACS data of the Virgo and field samples provide {\it uniform} multi-wavelength information on the frequency of SMBHs and NSCs in the local universe, across the mass spectrum and across environment.  In this work, we report on the analysis of the ACS observations of the field targets.  We model the galaxies' surface brightness profiles to determine what fraction host NSCs, and compare our results to those from the Virgo cluster. 
This paper is organized as follows.  Section 2 describes our sample, data reduction, and analysis.  Section 3 presents our results on the fraction of nucleated (as in hosting a NSC), early-type field galaxies, and our comparison to the fraction found for the Virgo Cluster.  Section 4 discusses the implications of these results in the context of NSC formation models.

\section{Data Analysis}

The full AMUSE-Field sample is comprised of 103 early-type galaxies. We refer the reader to Section 2 of \cite{:zr} for a detailed description of the selection criteria.  Our HST program aimed at acquiring dual band images for the galaxies with detected nuclear X-ray emission from \cxo\ (52 out 103 objects). Out of those, 8 already had archival  HST data in both the F475W and F850LP filters. The remaining 44 targets were approved for a Snapshot survey in Cycle 19 (PI: Gallo, ID 12951), and 17 of them were eventually observed, with 340.0 s and 340.0$+$60.0 s exposures for the F475W and F850LP filters, respectively (the completion rate was close to 40\%, in line with the average value for HST Snapshot programs).  For the purpose of estimating the nucleation fraction -- regardless of the nuclear X-ray properties -- we included 3 additional objects from the AMUSE-Field sample with no X-ray detection, but for which ACS data in the same filters was also available in the archive.  
To summarize, here we report on HST/ACS observations in the F475W \& F850LP filters for 28 (17 new, plus 11 archival) out of the 103 AMUSE-Field galaxies; 25 out of those 28 belong to the subsample of 52 galaxies with X-ray detected nuclei. 

Data was re-reduced using the AstroDrizzle pipeline to improve the point spread function sampling from 0\farcs05/pixel to 0\farcs03/pixel for all objects.  
Though all objects were originally classified as early-type in HyperLeda, three objects (ESO 540-014, NGC 0855, NGC 3265) revealed irregular or non-early type morphology in the ACS images, and were thus excluded from further analysis.  We refer to the Appendix for descriptions of individual objects. 
We corrected for extinction within our galaxy using the extinction maps by \cite{1998ApJ...500..525S}.  Additionally, seven objects (NGC 1172, NGC 1370, NGC 3073, NGC 3377, NGC 4036, NGC 4125, and NGC 4278) were identified as suffering from dust contamination through visual inspection.  We corrected for dust following the procedure defined in \cite{:fs}, which relies on interpolation across dust affected areas.  Masses for both galaxies and NSCs were computed using the mass-to-light relations for Sloan \textit{g} and \textit{z} bands\footnote{Sloan \textit{g} and \textit{z} bands are roughly equivalent to HST F475W and F850LP bands, respectively.} defined in Table 7 of Appendix A of \cite{2003ApJS..149..289B}.  In calculating masses, magnitude corrections were made for the slight deviations between the HST F475W \& F850LP and Sloan \textit{g} \& \textit{z} filters (Eric Bell, private communication). 
Excluding the three aforementioned non-early type galaxies and two galaxies (NGC 1370 and NGC 3073) for which heavy dust contamination around the nucleus prevented us from doing accurate photometry, our analysis is based on a sample of 23 early-type, field galaxies.  Properties of individual galaxies are listed in Table \ref{OptProps}. 

\begin{figure*}
\centering
\begin{tabular}{lcr}
\includegraphics[scale=0.22]{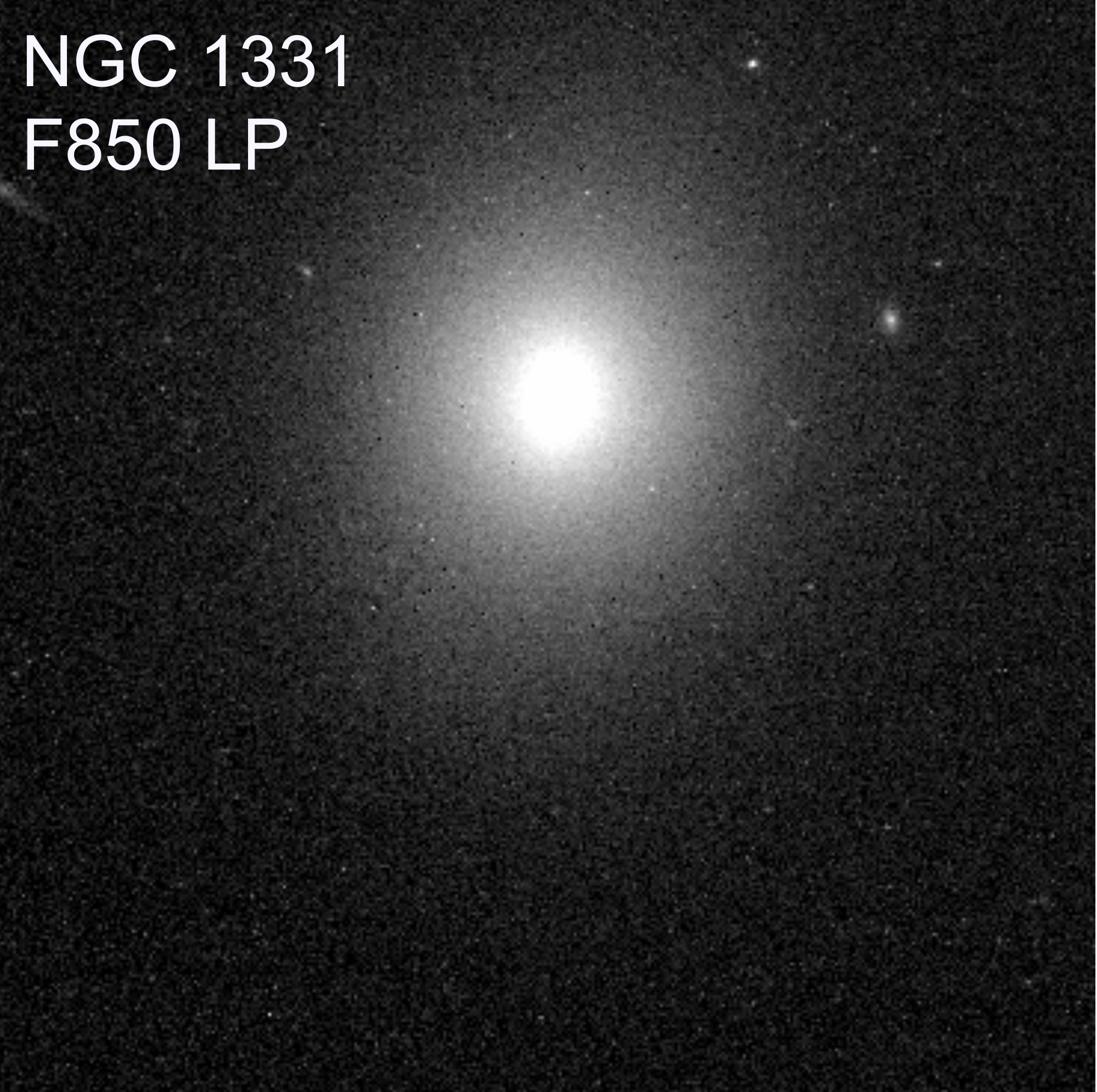}  &
\includegraphics[scale=0.22]{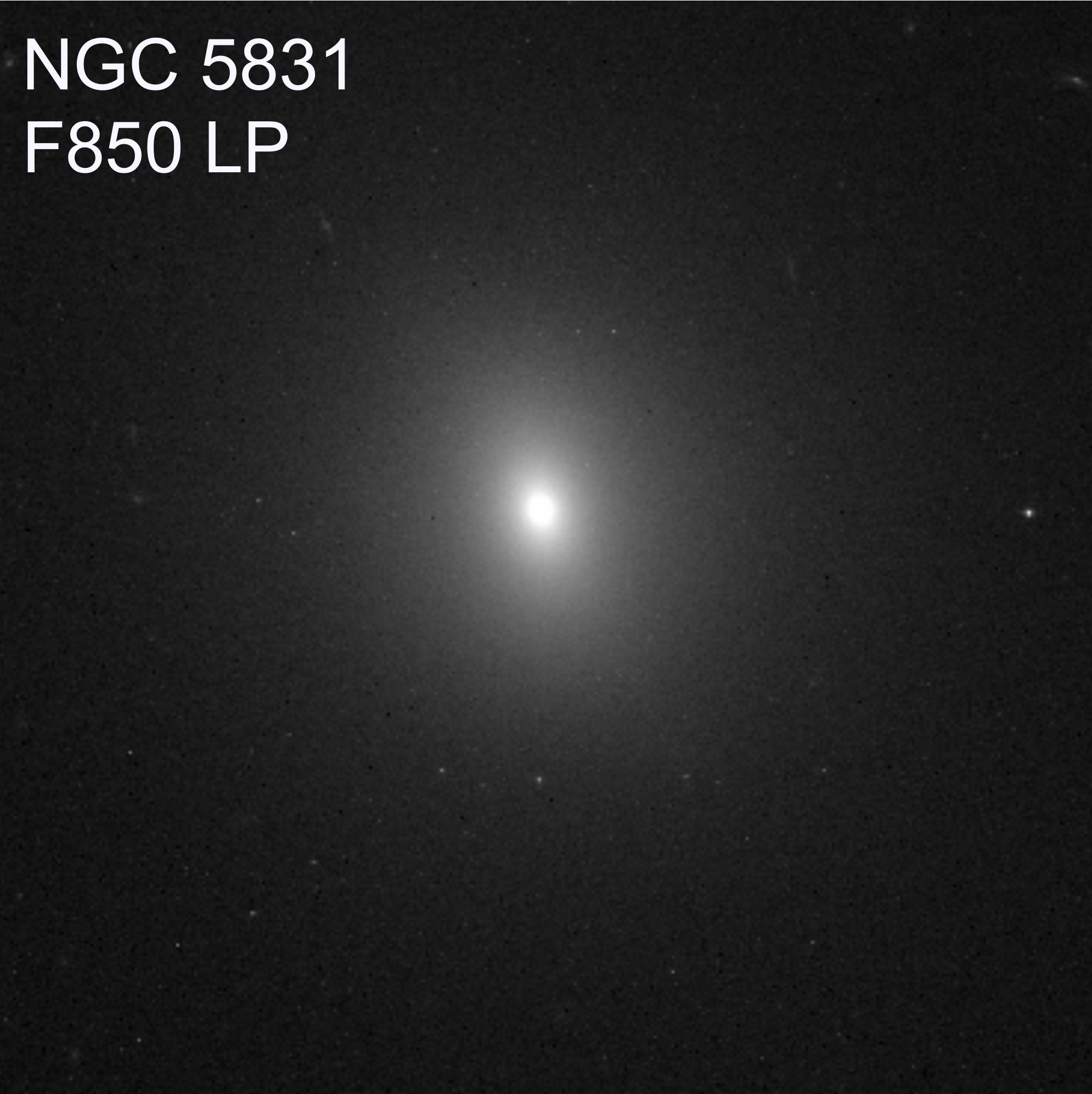}  &
\includegraphics[scale=0.22]{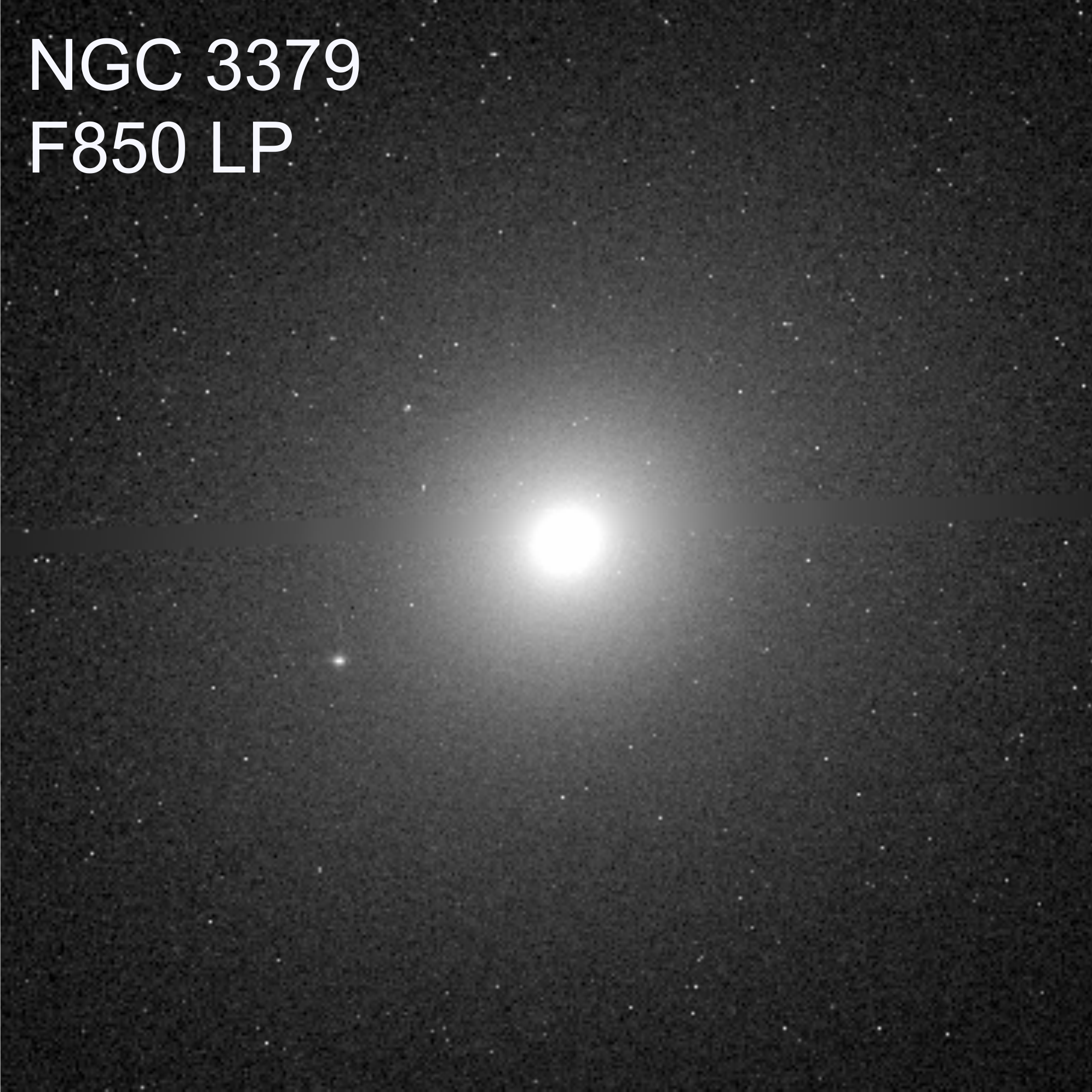} \\

\includegraphics[scale=0.22]{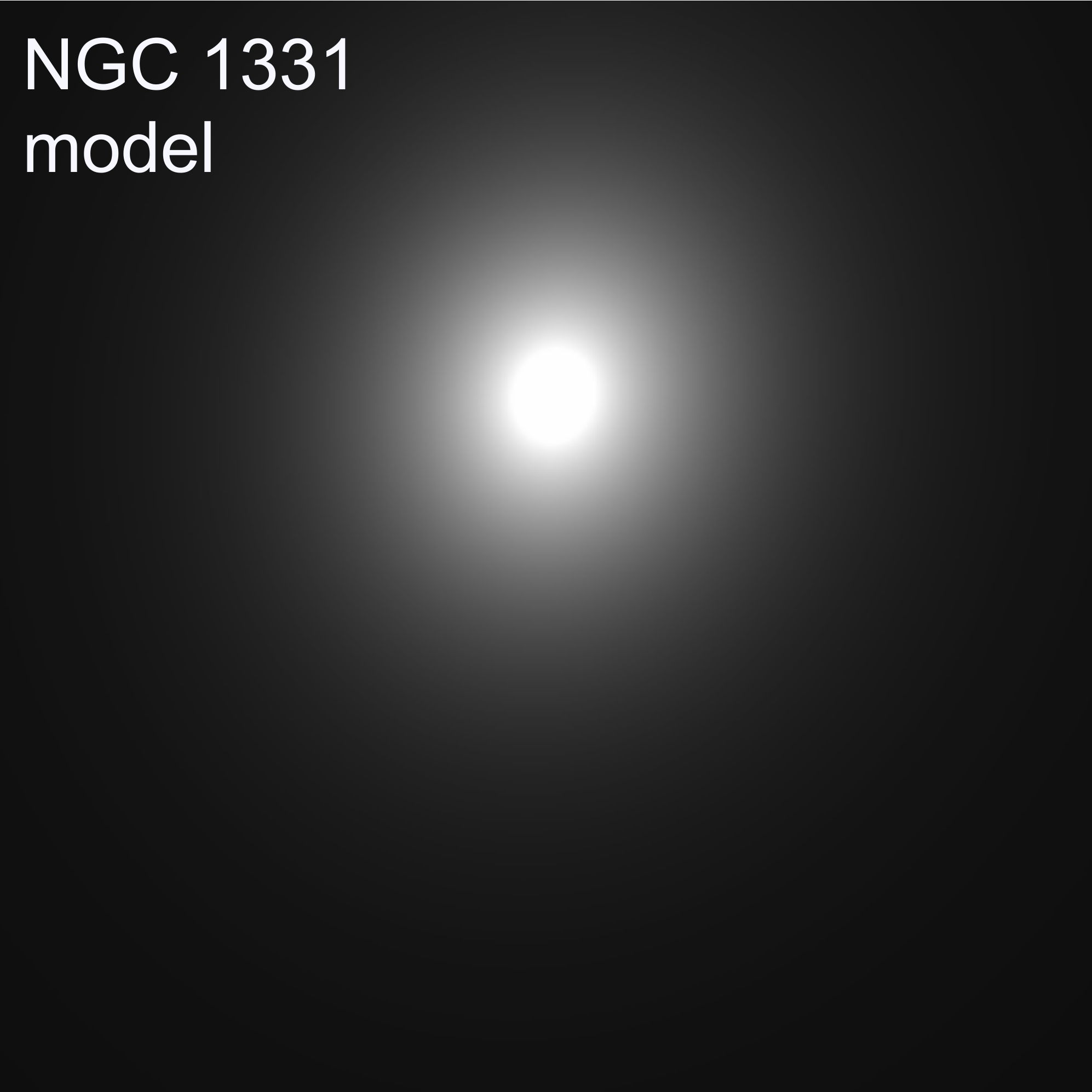} &
\includegraphics[scale=0.22]{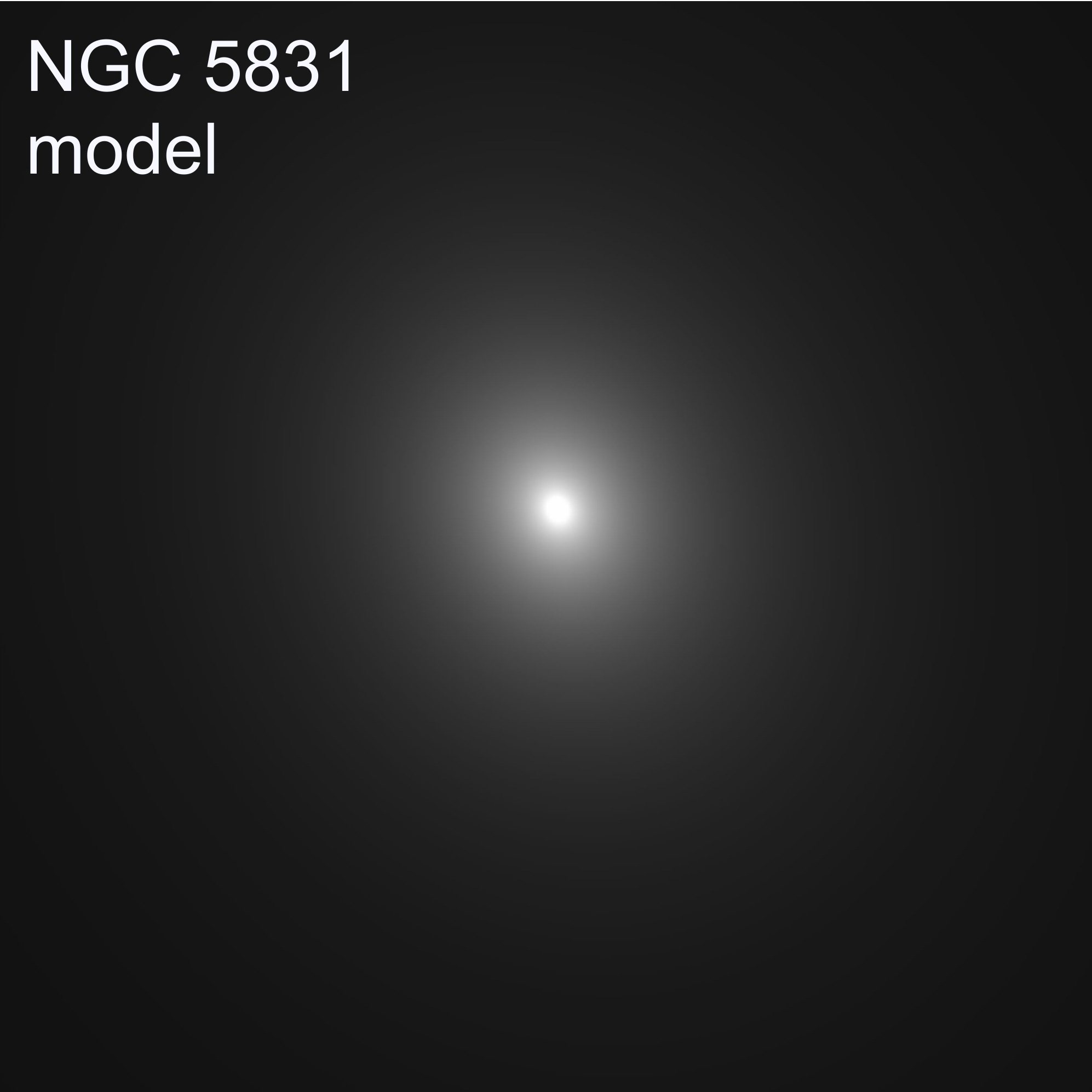} &
\includegraphics[scale=0.22]{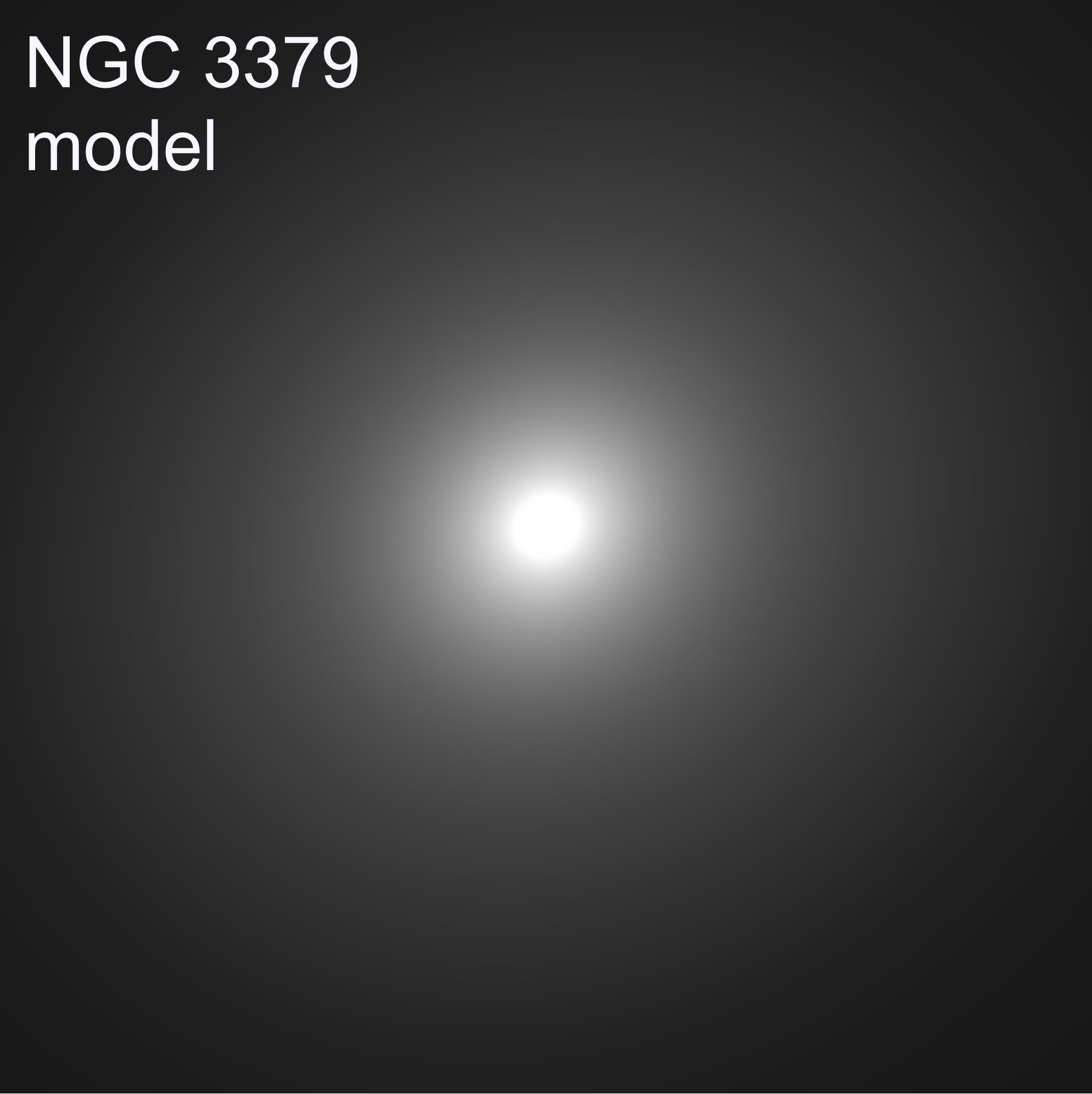} \\

\includegraphics[scale=0.22]{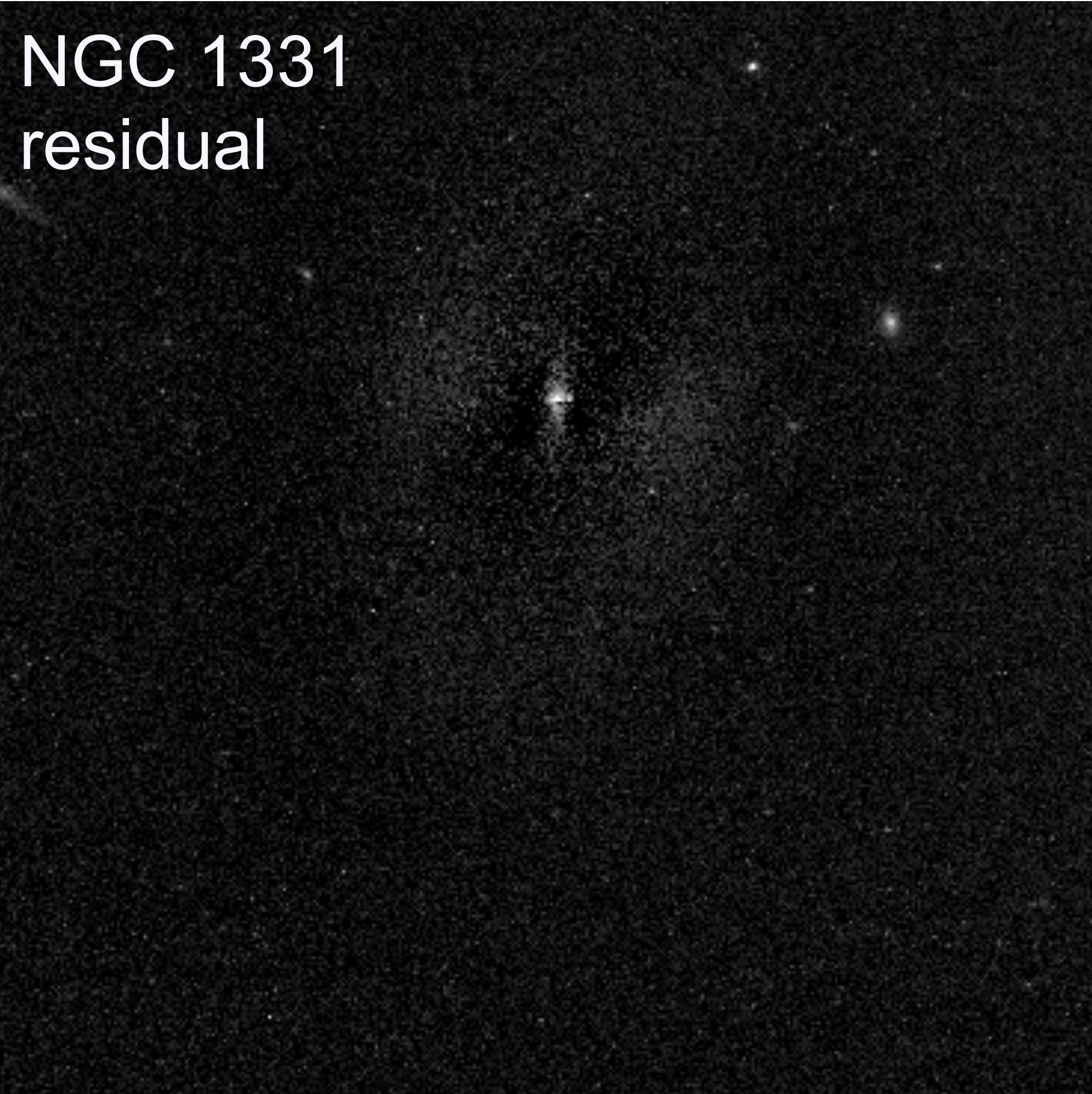} &
\includegraphics[scale=0.22]{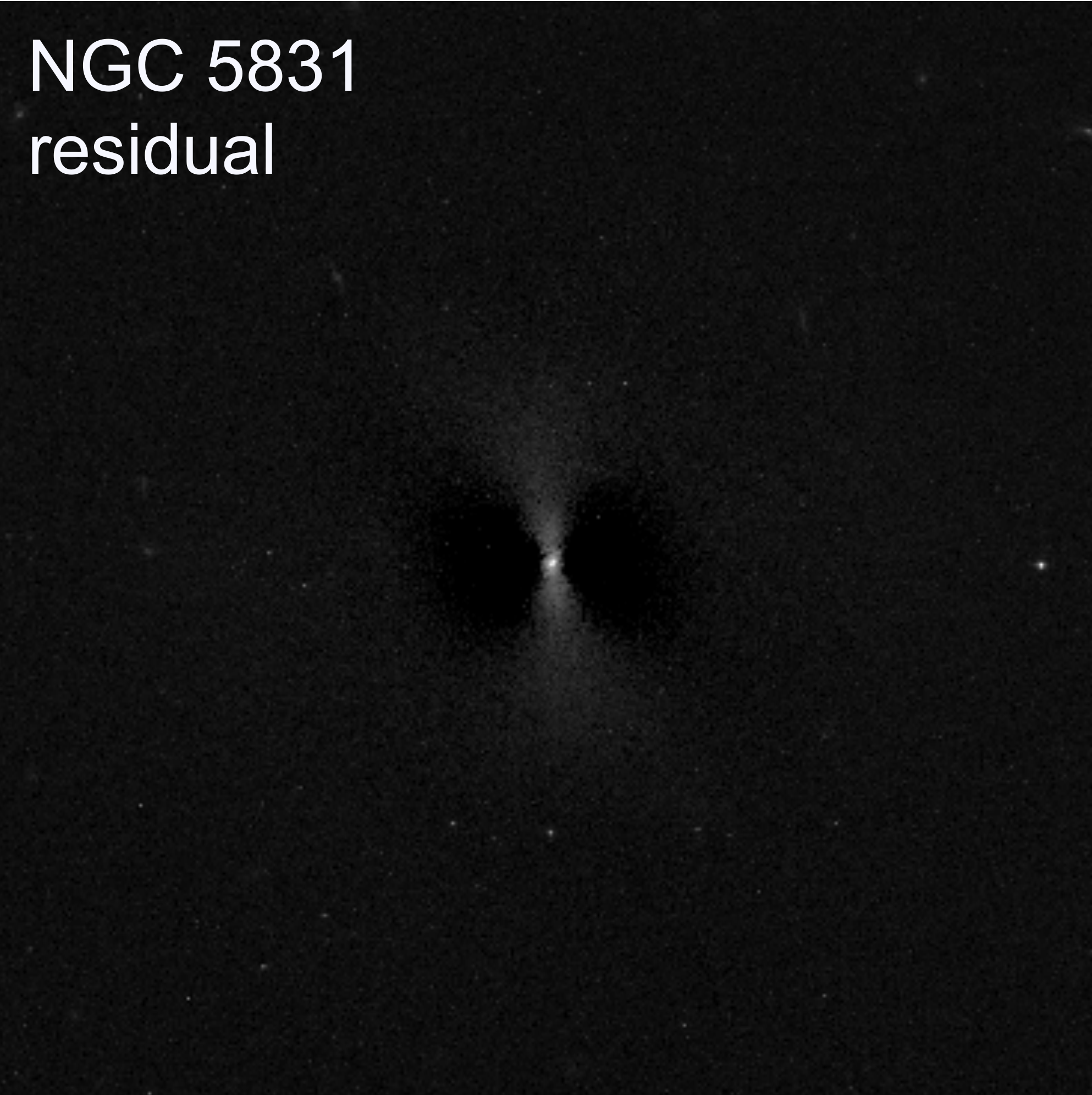}  &
\includegraphics[scale=0.22]{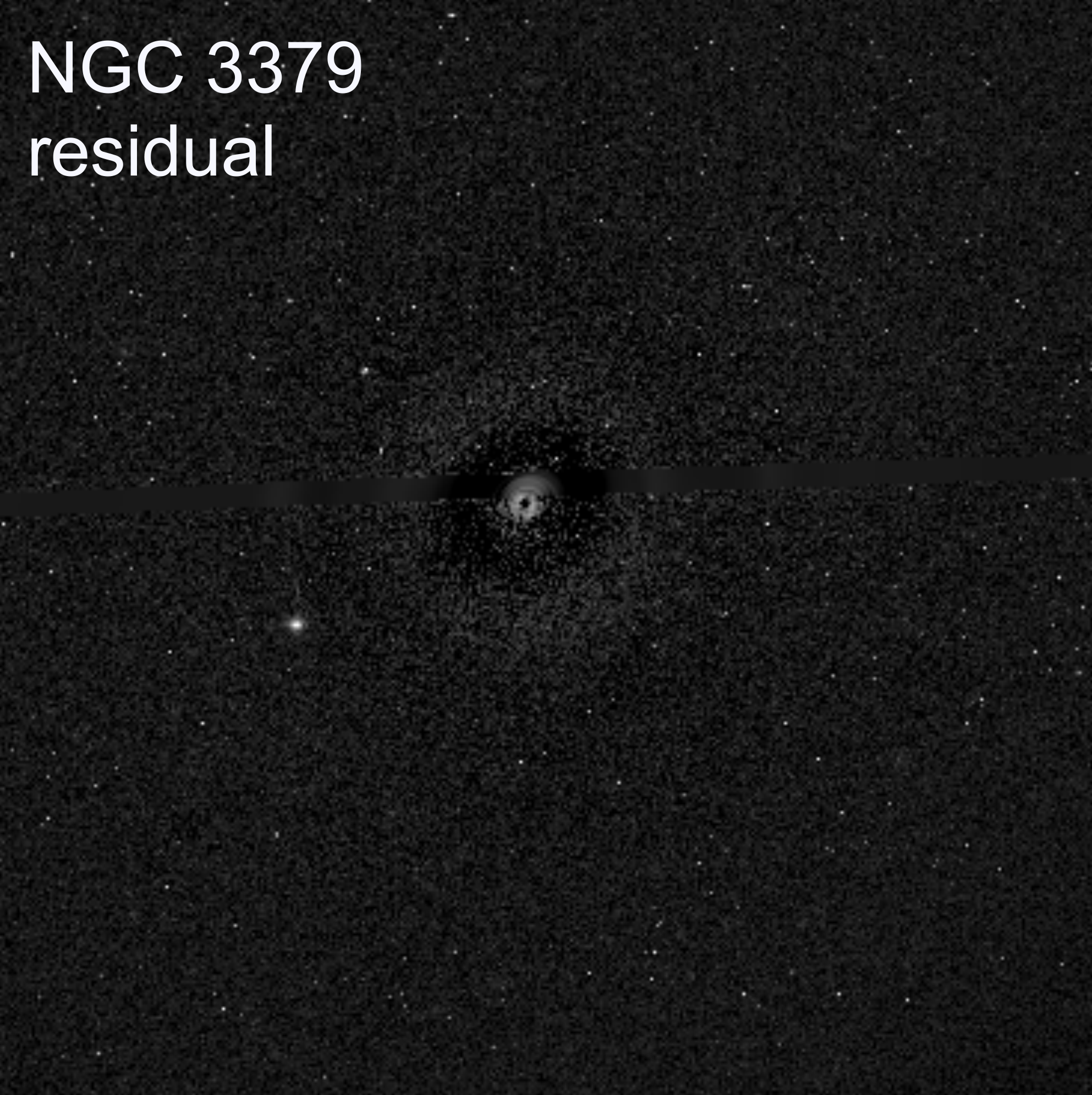} \\

\end{tabular}
\caption{HST/ACS F850LP images, GALFIT 2-D models, and GALFIT residuals for three AMUSE-Field targets.  NGC 1331 (left column) was fit with a double-S\'{e}rsic profile; NGC 5831 (center column) was fit with a single S\'{e}rsic profile; NGC 3379 was fit with a S\'{e}rsic profile and an additional outer, low-S\'{e}rsic index component.  Scaling is the same within each image/model/residual set.}
\label{models}
\end{figure*}

For the purpose of a meaningful comparison to the Virgo Cluster nucleation fraction, we follow the definition of nucleation given by \cite{2006ApJS..165...57C}, whereby a nucleated galaxy is one whose inner light profile lies systematically above that of a S\'{e}rsic profile\footnote{A S\'{e}rsic profile is defined as  
\begin{equation}
I_{\rm S\acute{e}rsic}(R) = I_{\rm e} \exp\left\{ -b_{\rm n} \left[ \left( \frac{R}{R_{\rm e}} \right)^{1/n} - 1 \right] \right\} \,
\end{equation}
where $R_{\rm e}$ is the effective radius which encloses half of the model's light, $I_{\rm e}$ is the intensity at the effective radius, $n$ is the S\'{e}rsic index, and $b_{\rm n}$ is a constant that is dependent on $n$.  } fit to the galaxy excluding the inner 0\farcs5. 
Surface brightness profiles were parametrized with a S\'{e}rsic or double-S\'{e}rsic profile.  We used the IRAF package ELLIPSE to extract azimuthally averaged surface brightness profiles from both the F475W and F850LP images.  Two dimensional galaxy models were created with GALFIT \citep{2002AJ....124..266P} and then projected onto the 1-D surface brightness profiles.
First, we attempted to fit a S\'{e}rsic profile to the overall galaxy, excluding the inner 0\farcs5 (as done in \citealt{2006ApJS..165...57C}; note that the Field sample was selected to have approximately the same distance as Virgo).   When needed, an exponential disk component was added to the model to fit the outer regions of the galaxy (typically beyond 50\arcsec, i.e. \citealt{2013ApJ...768...36D}).  While GALFIT does not allow the position angle and isophotal ellipticity to radially vary within a component, we did not constrain components to have the same position angle and ellipticity. Example GALFIT models and residuals are shown in Figure~\ref{models}.

The presence of a NSC is signaled by a light excess with respect to the extrapolation of the best-fit galaxy model within the inner 0\farcs5. When present, such excess was fit with an additional S\'{e}rsic component.  A good fit was chosen to have residuals between the data and model consistently between $-$0.2 and $+$0.2 mags (similar to the residuals for surface brightness profile fits of \citealt{2006ApJS..165...57C}). Of the 23 objects for which we carried out a surface brightness profile analysis, we find six of them to be nucleated based on the definition given by \cite{2006ApJS..165...57C}.  Table~\ref{NSC_props} lists half-light radii, masses, and colors of these six NSCs.  

\begin{figure*}
\centering
\begin{tabular}{cc}

\includegraphics[scale=0.23]{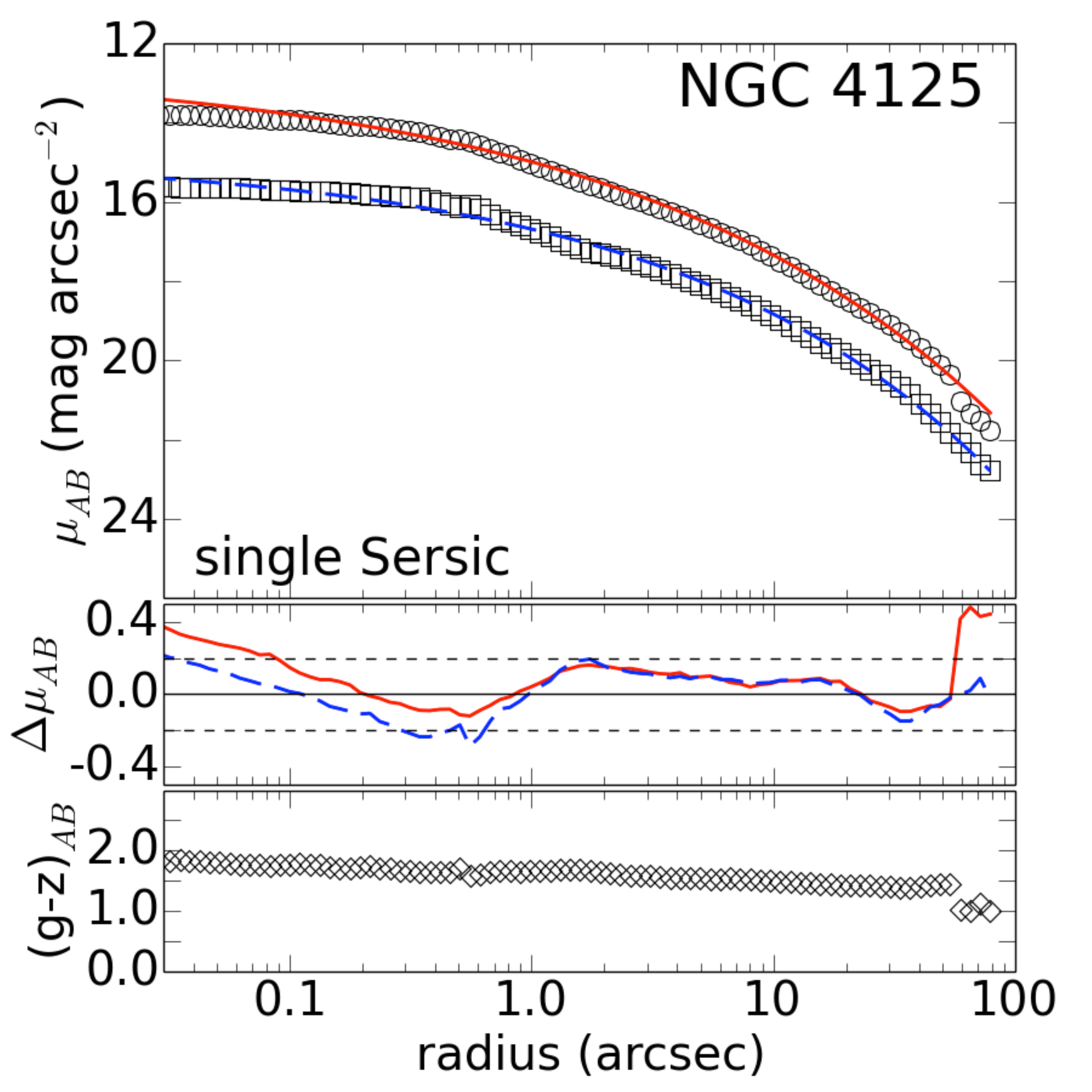} & \includegraphics[scale=0.23]{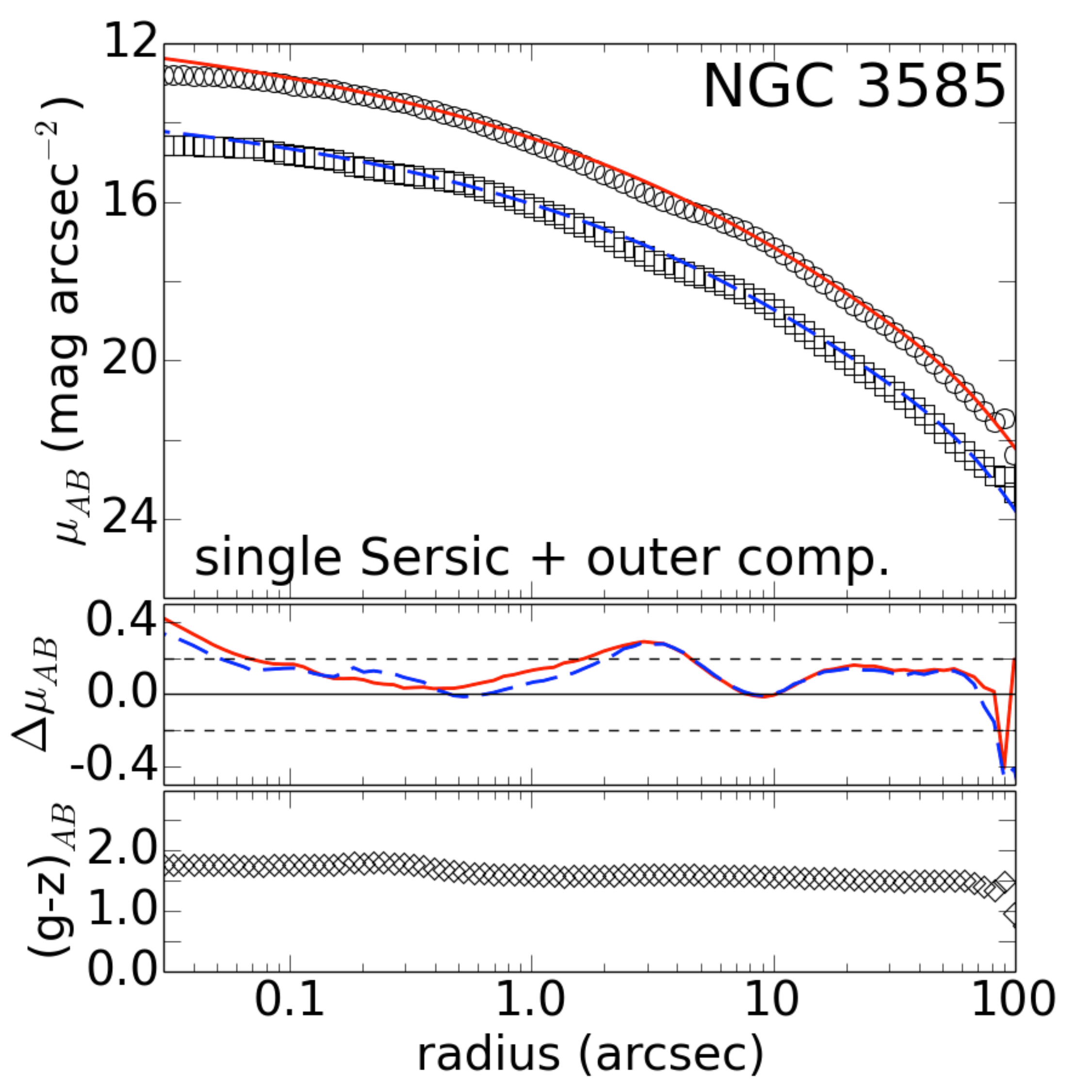} \\

\includegraphics[scale=0.23]{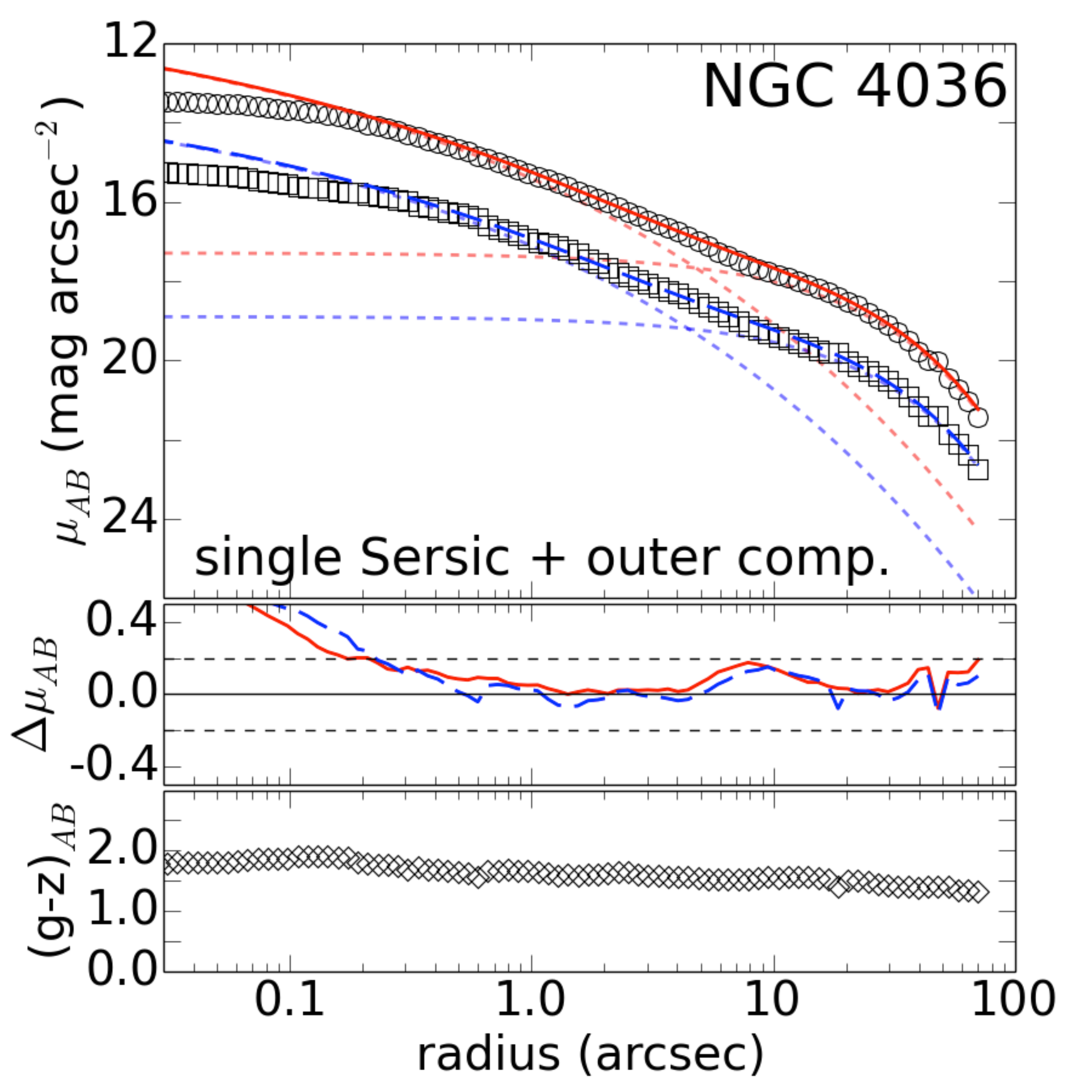} & \includegraphics[scale=0.23]{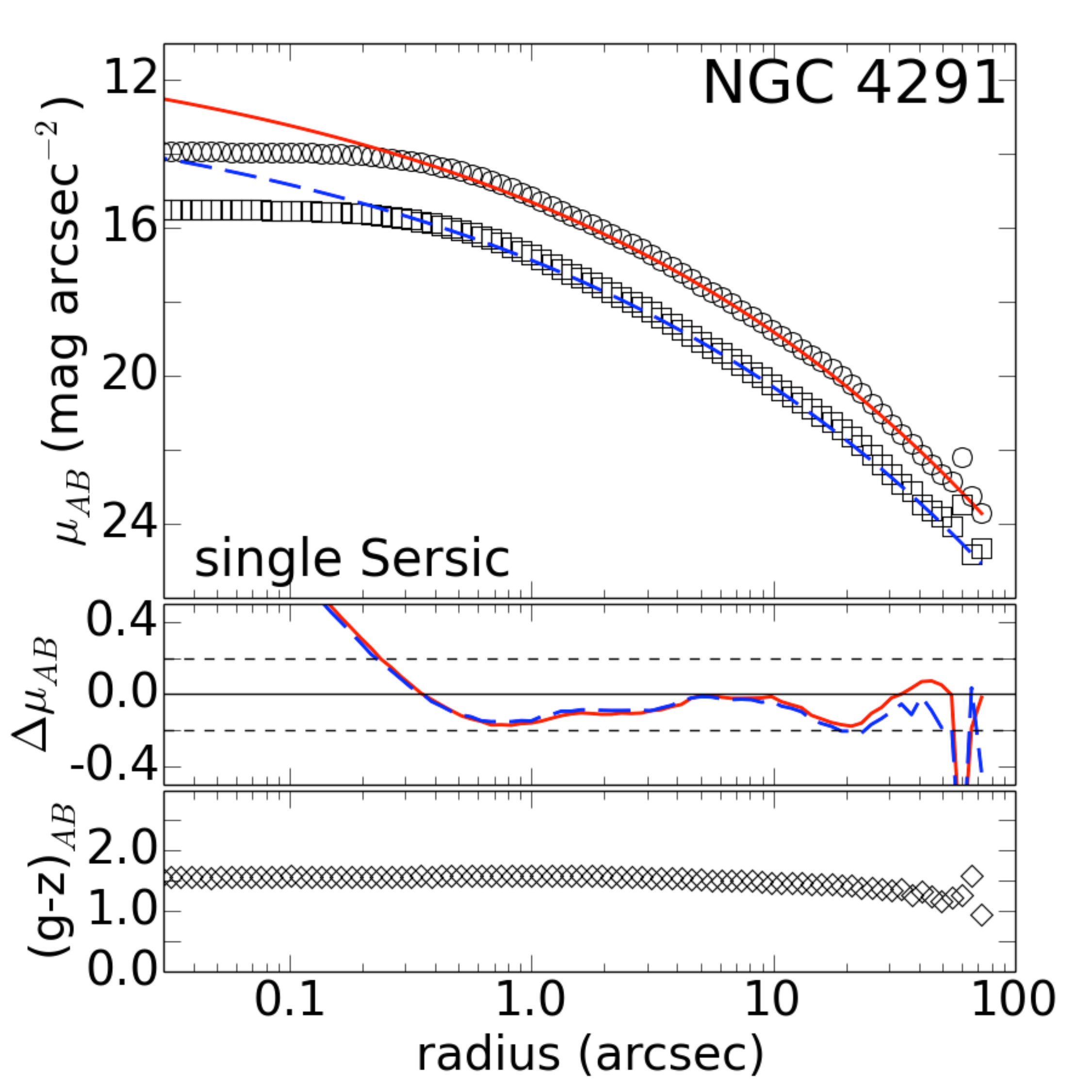} \\

\includegraphics[scale=0.23]{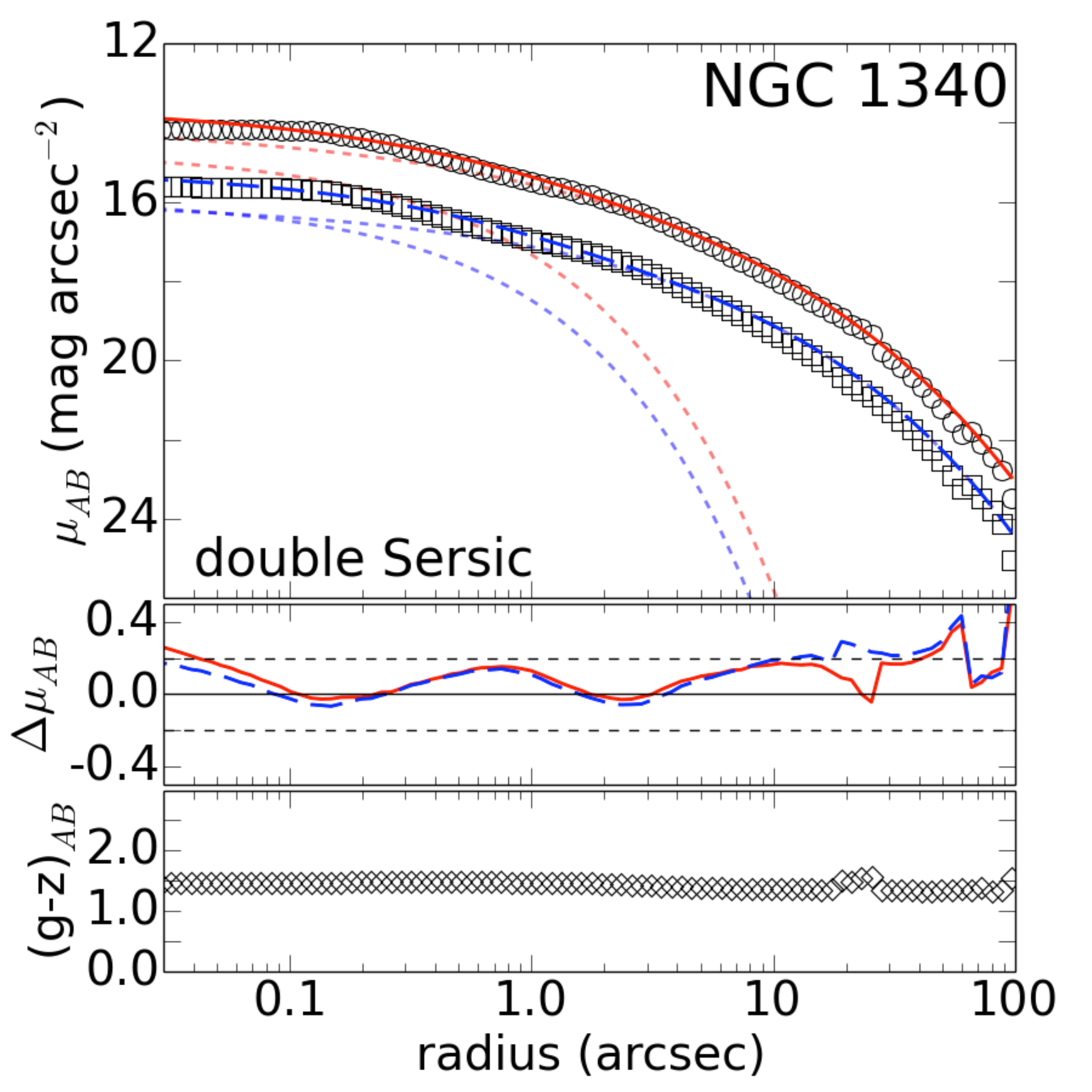} & \includegraphics[scale=0.23]{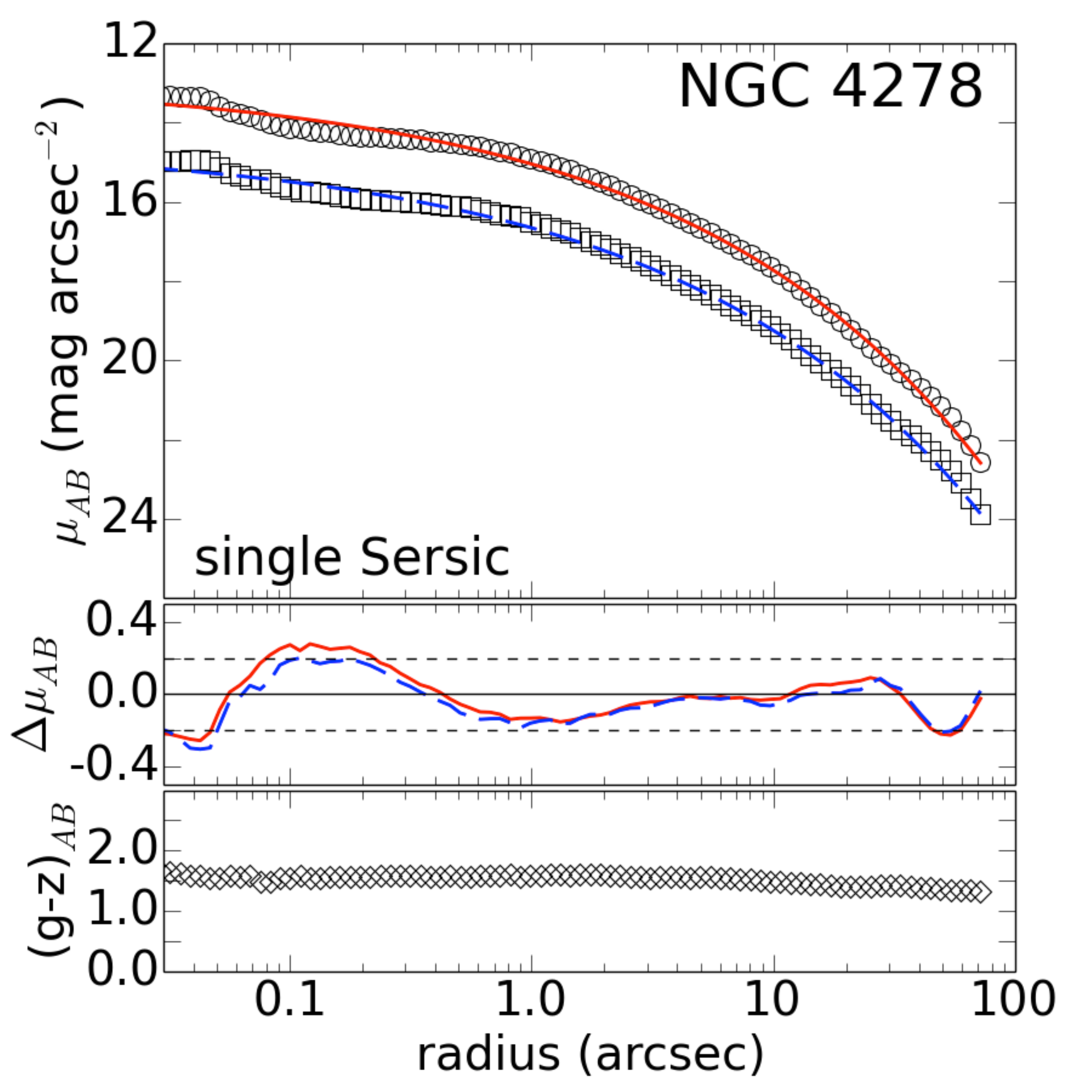} \\

\end{tabular}
\caption{For each AMUSE-Field object with HST coverage, we present a 1-D light profile, best fit GALFIT model profile (projected to a 1-D surface brightness profile), residuals, and color gradient.  The semi-major axis returned by ELLIPSE is plotted on the x-axis. The top panel of each plot gives the light profile and the best fit model profile (a single or double S\'{e}rsic profile, either with or without an outer disk component).  Open circles represent the z band data, and open squares represent the g band data.  The best fit to the z band data is given by the solid, red line, while the best fit to the g band data is given by the blue, long dashed line.  If multiple components are necessary to fit a light profile, the individual components are shown in short dashed lines.  In the middle panel is the residual between the data and best fit profile (dashed, blue line for g band; solid, red line for z band).  The bottom panel is the (g-z) color gradient.}
\label{lightpro1}
\end{figure*}

\begin{figure*}
\centering
\begin{tabular}{cc}

\includegraphics[scale=0.23]{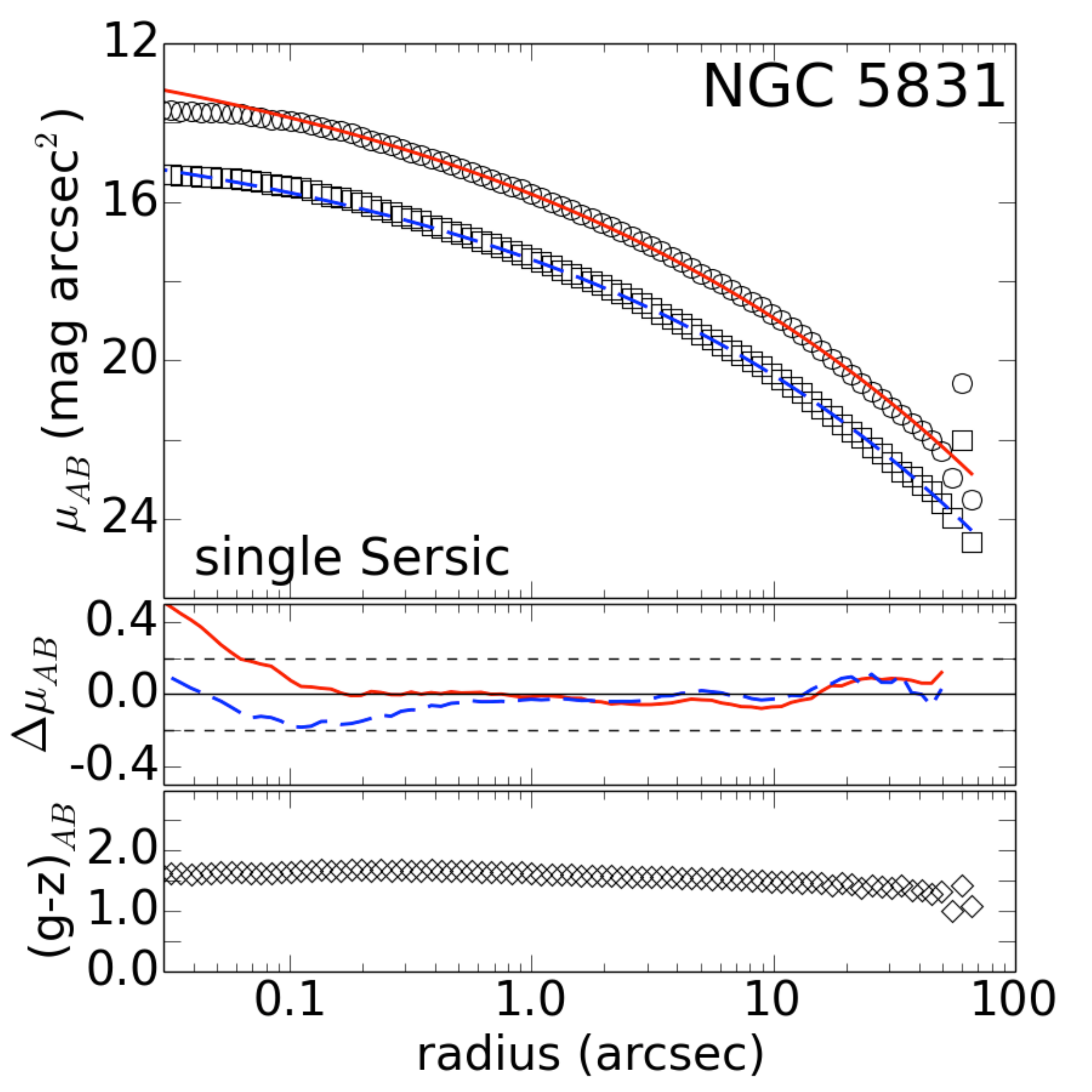} & \includegraphics[scale=0.23]{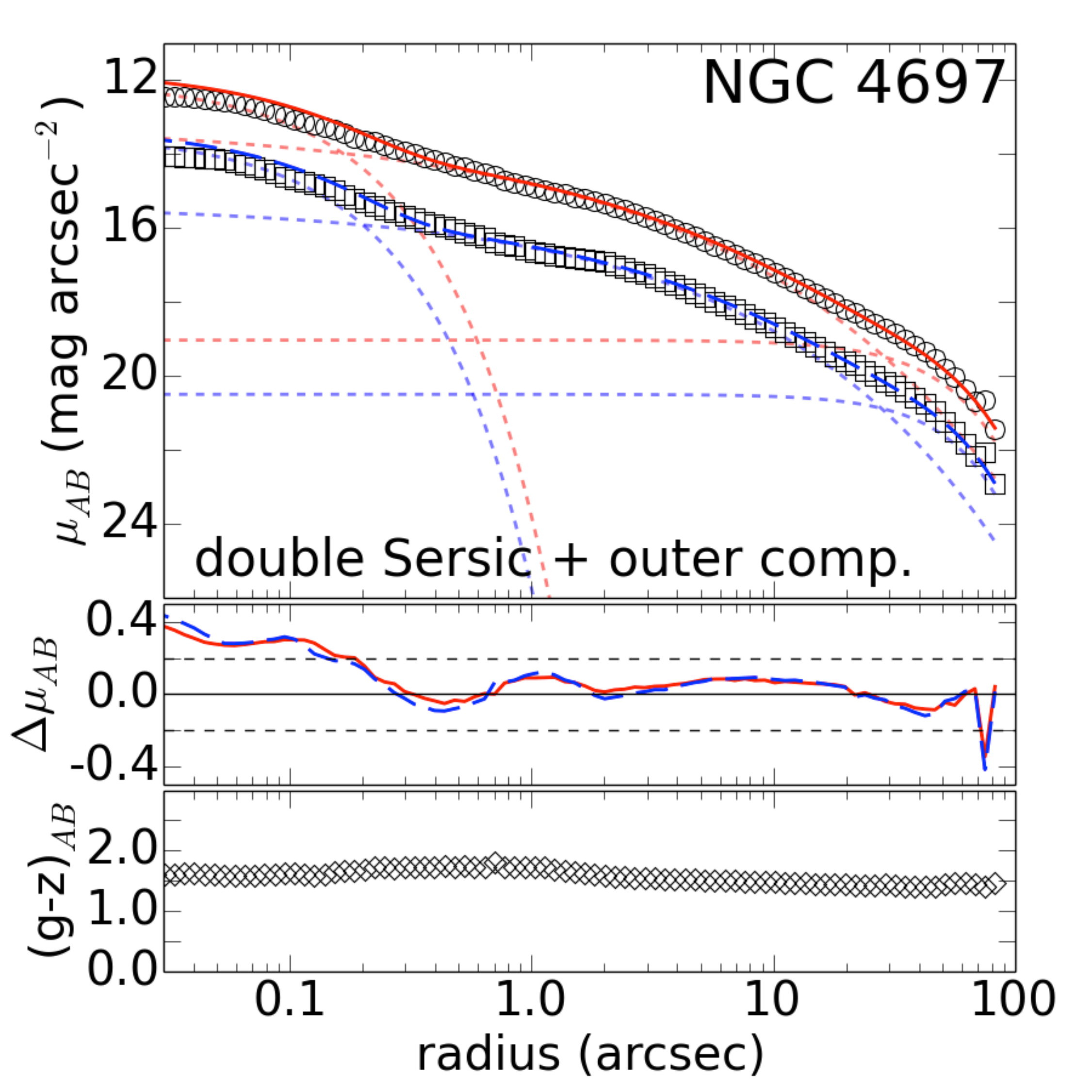} \\

\includegraphics[scale=0.23]{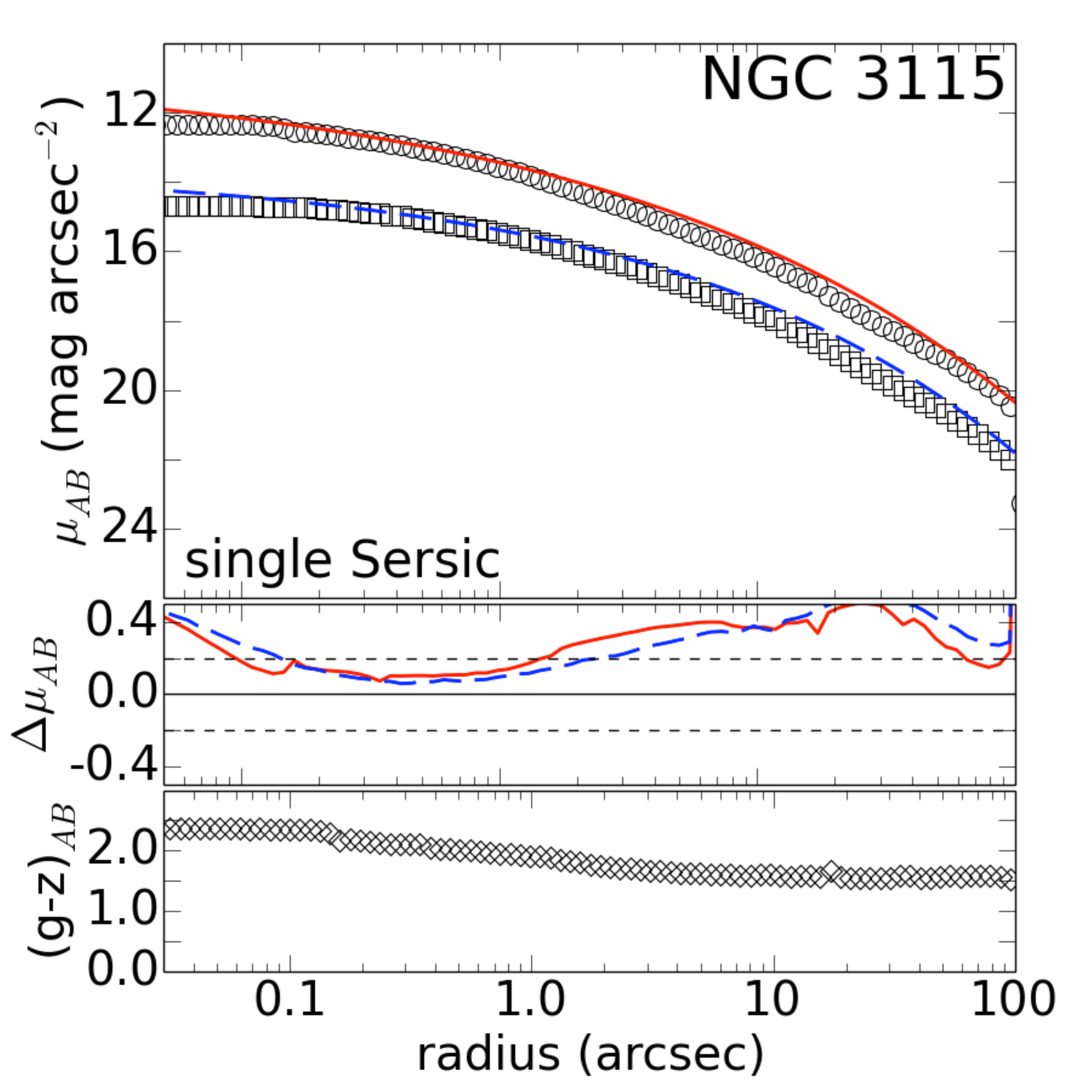} & \includegraphics[scale=0.23]{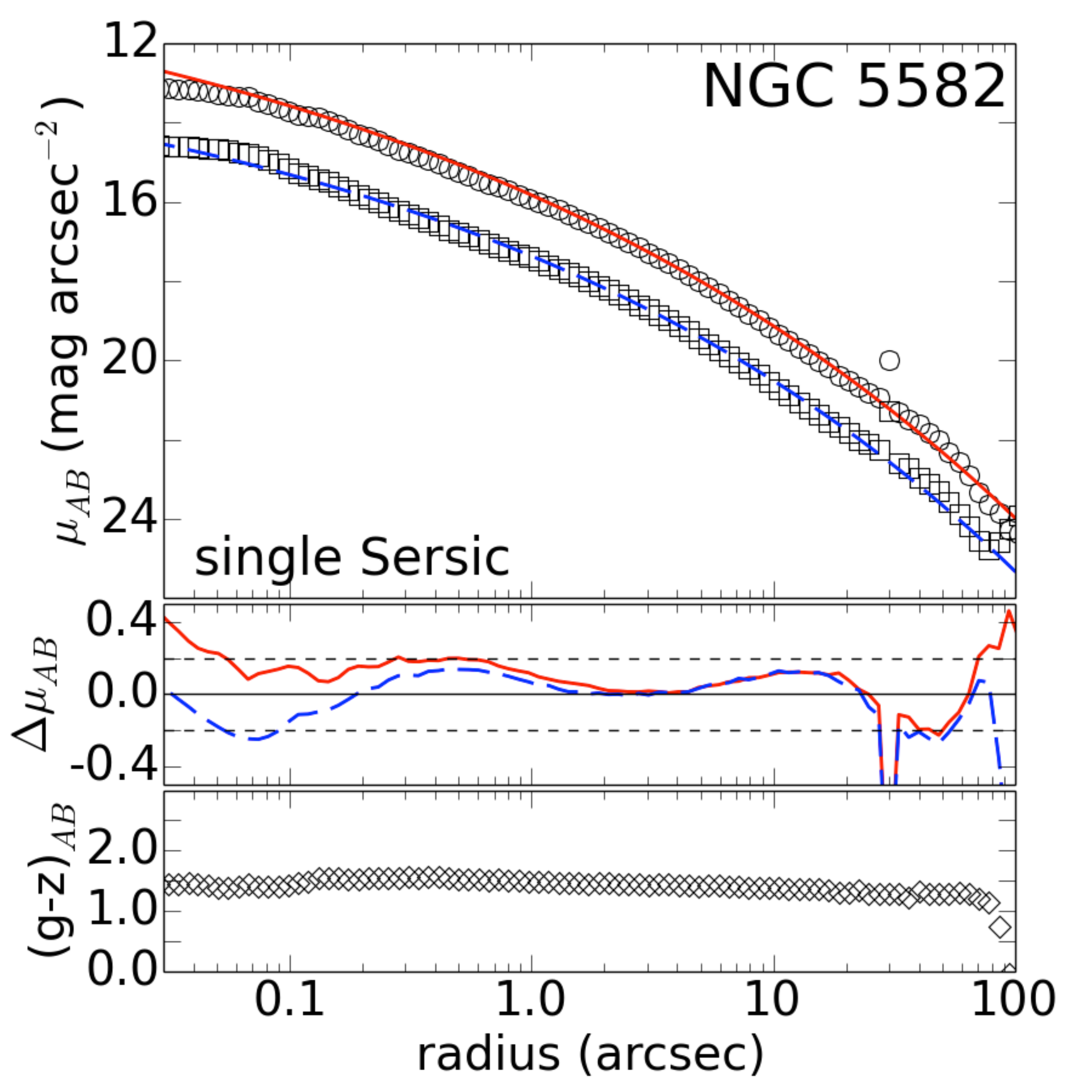} \\

\includegraphics[scale=0.23]{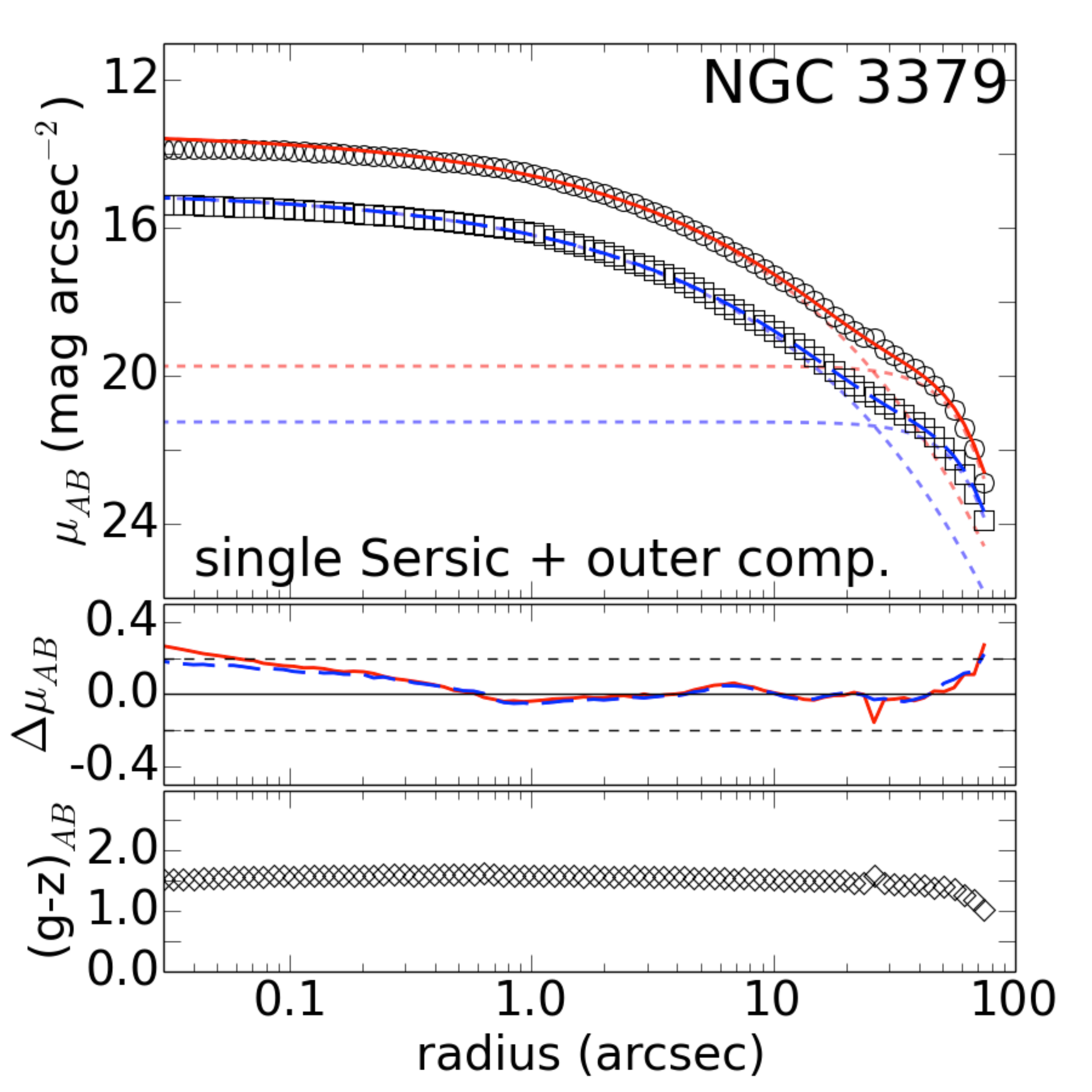} & \includegraphics[scale=0.23]{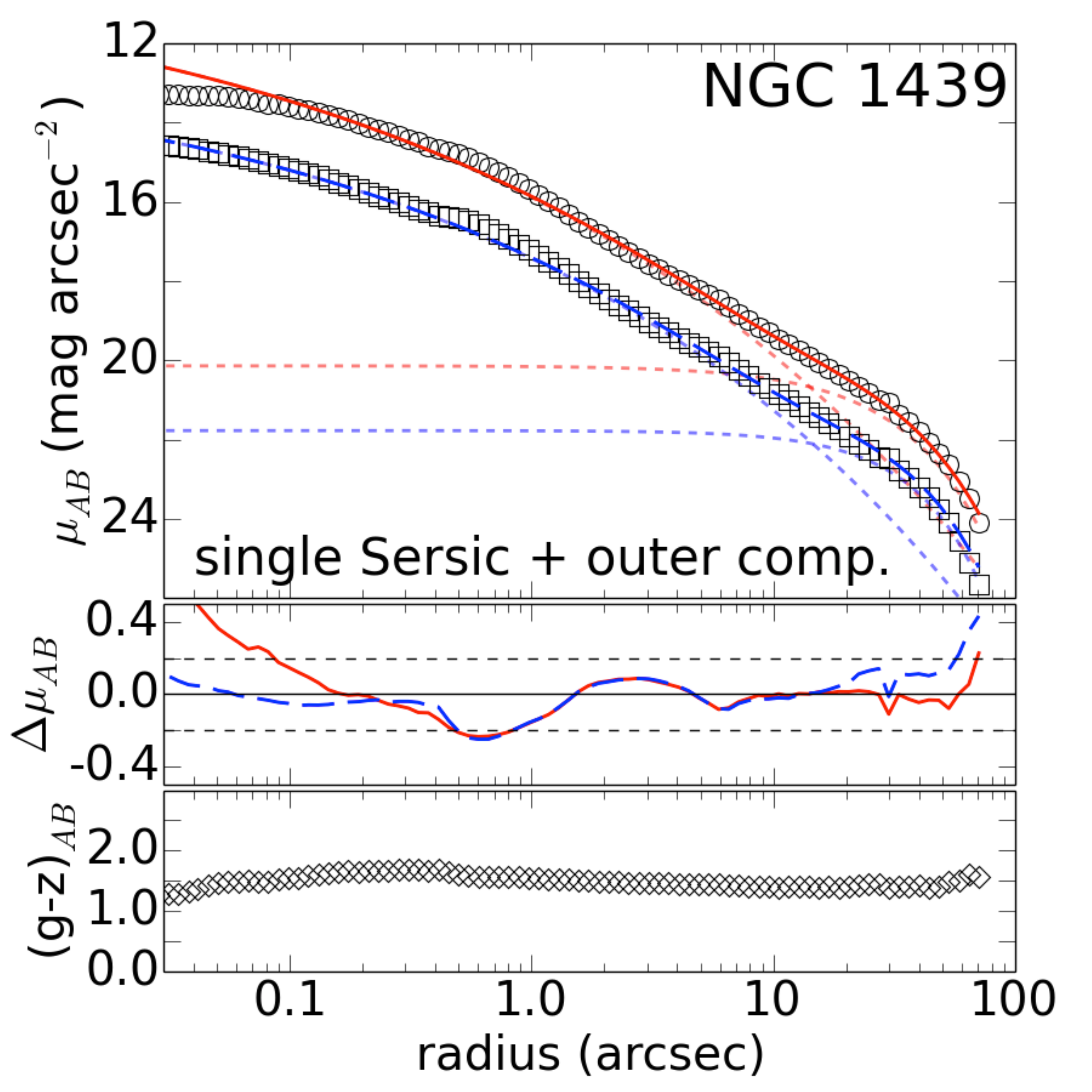}

\end{tabular}
\caption{Light profiles for AMUSE-Field objects with HST coverage.  See Figure~\ref{lightpro1} for description.}
\label{lightpro2}
\end{figure*}

\begin{figure*}
\centering
\begin{tabular}{cc}

\includegraphics[scale=0.23]{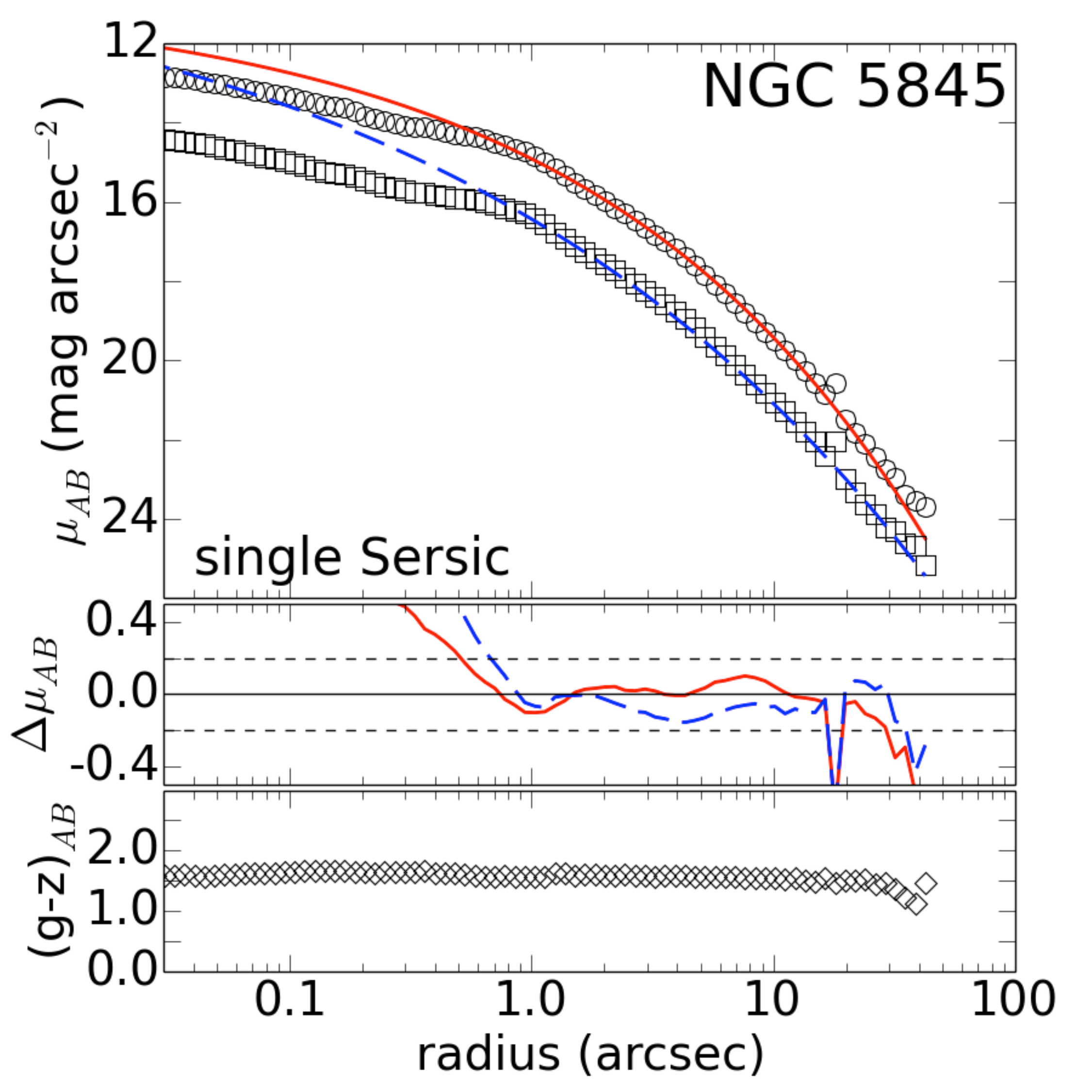} & \includegraphics[scale=0.23]{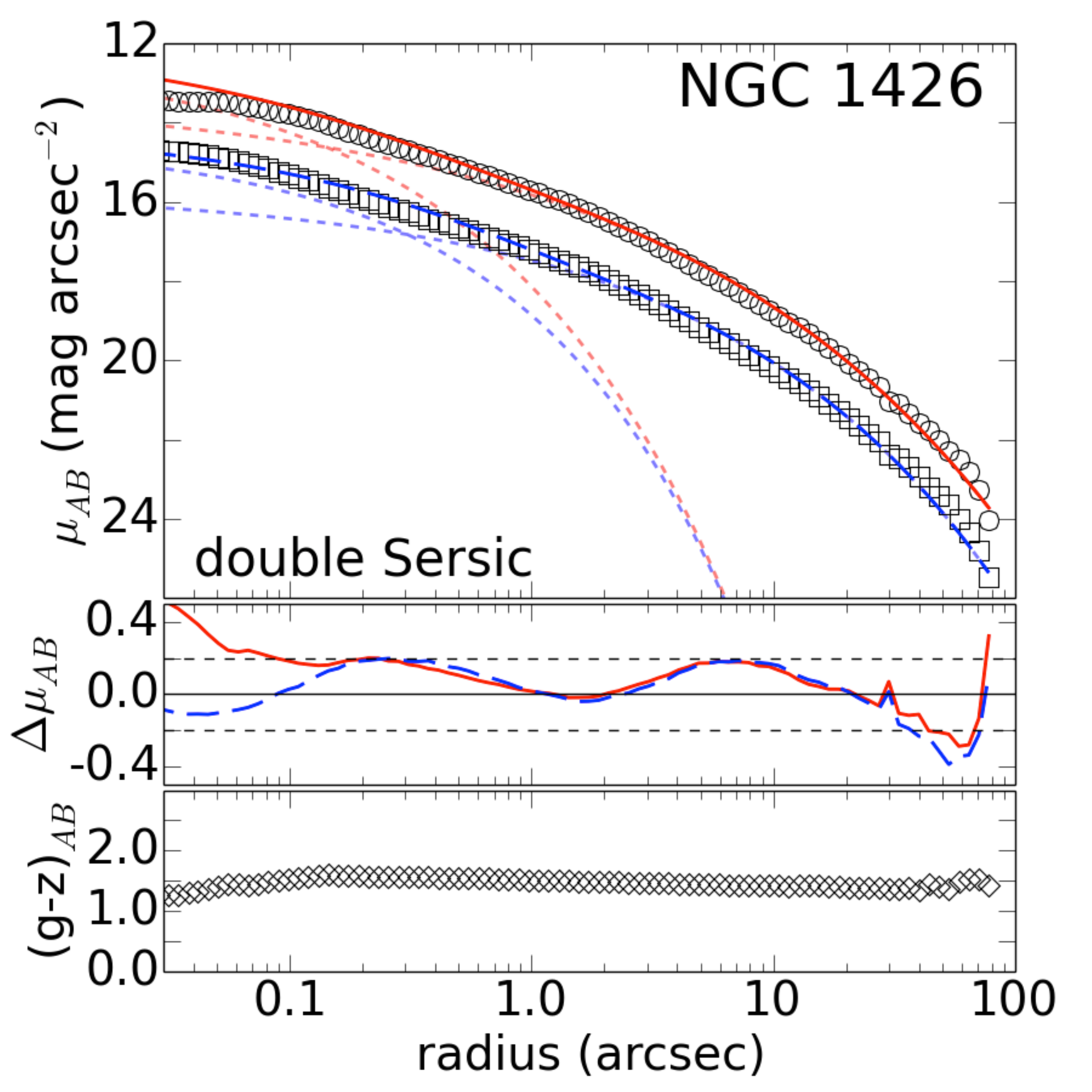} \\

\includegraphics[scale=0.23]{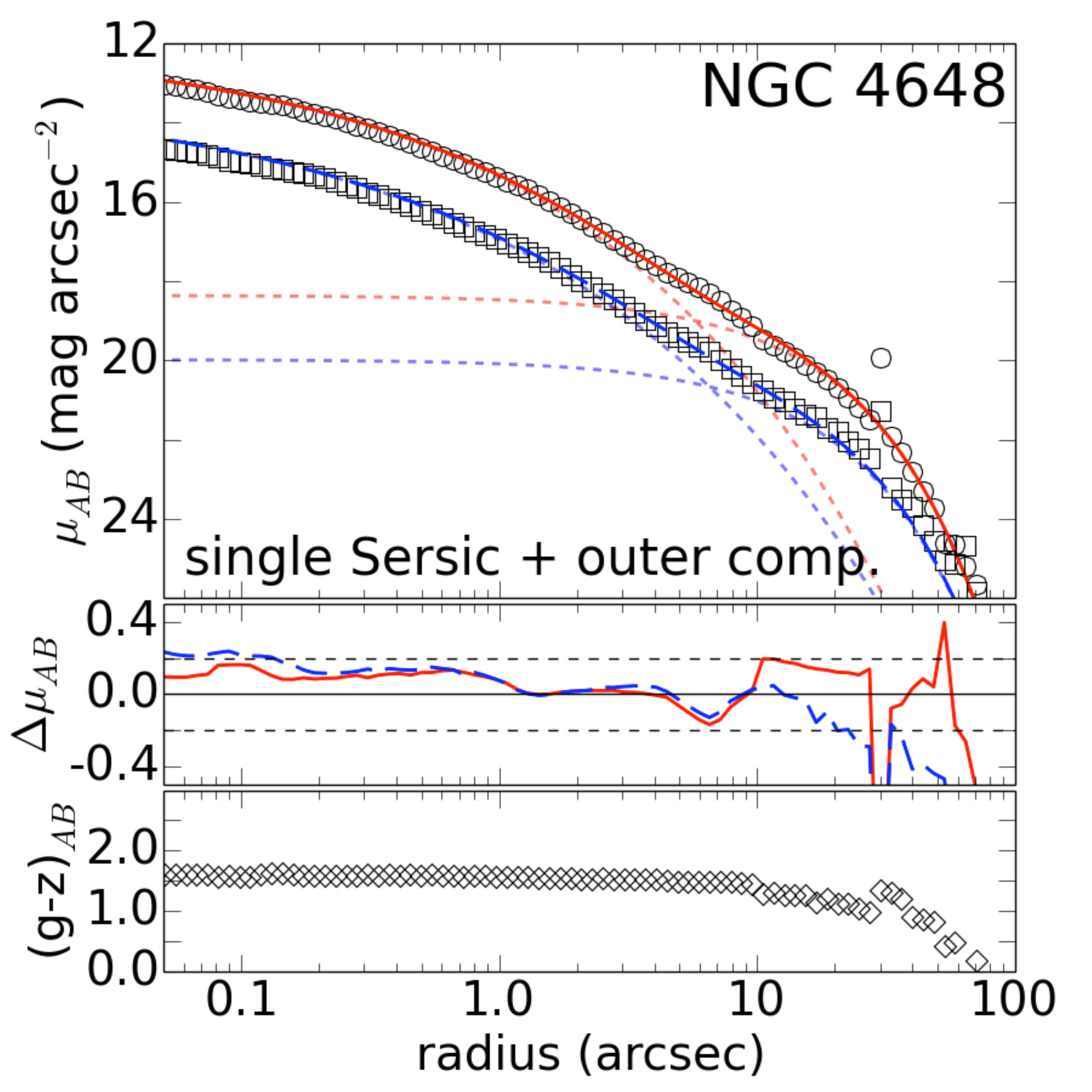} & \includegraphics[scale=0.23]{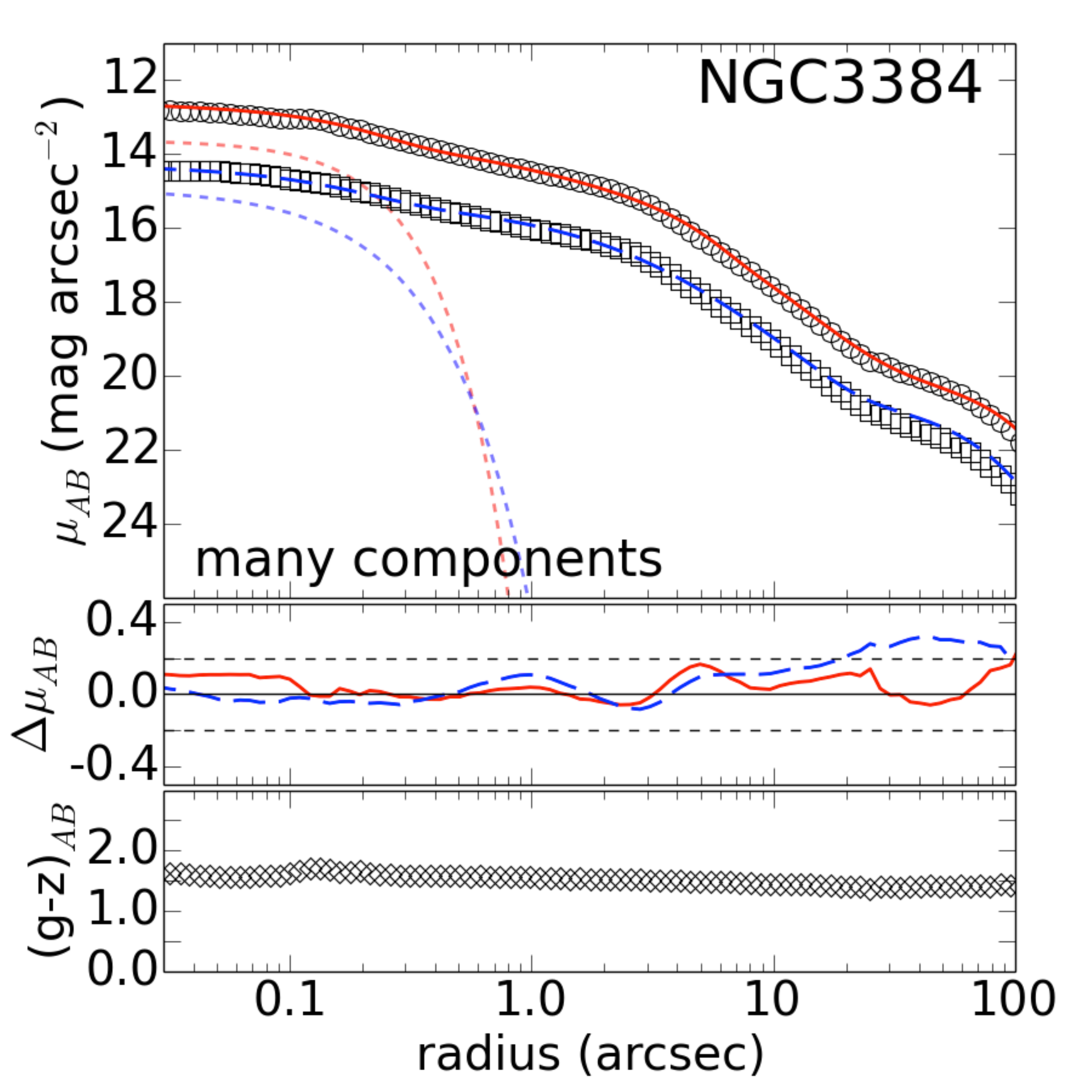} \\

\includegraphics[scale=0.23]{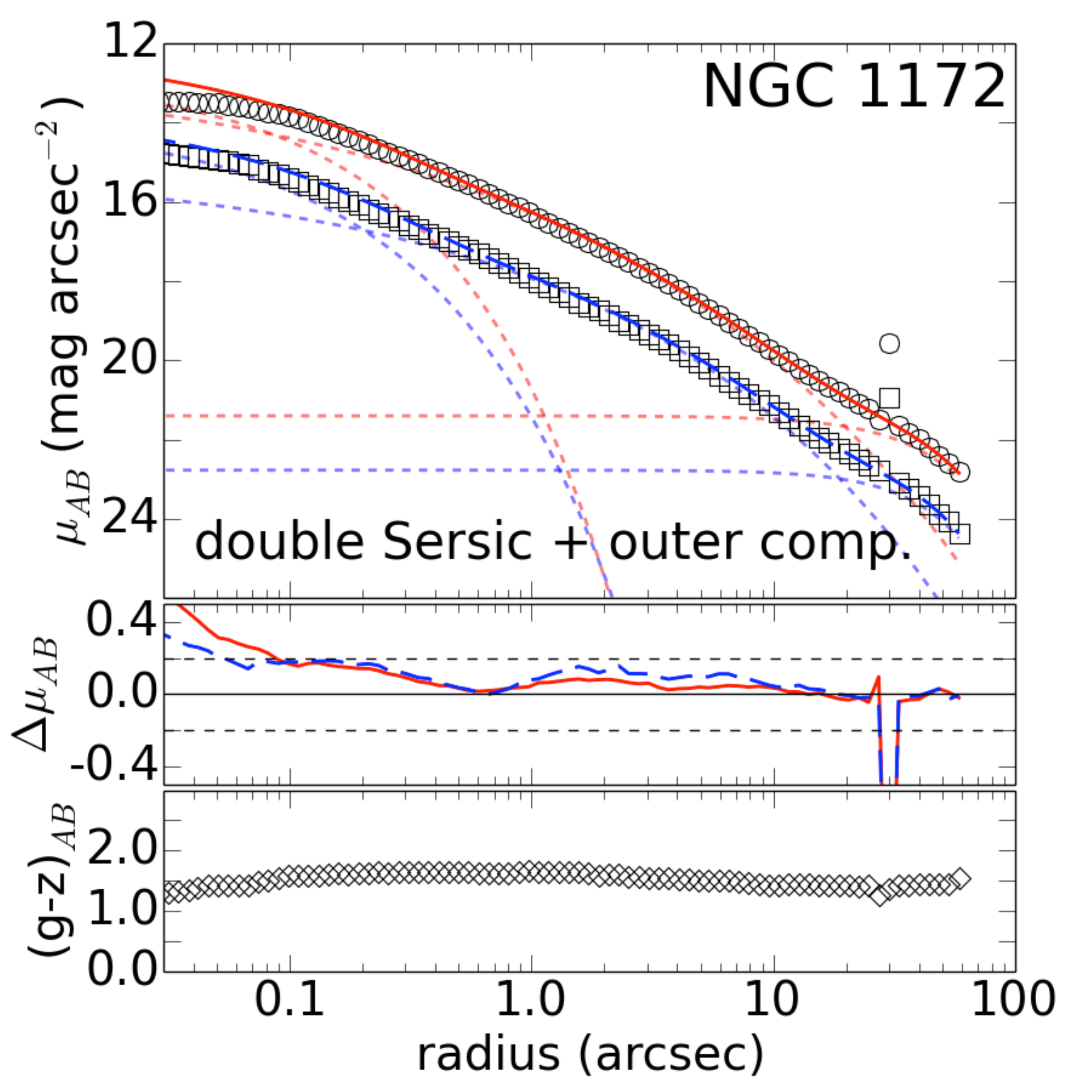} & \includegraphics[scale=0.23]{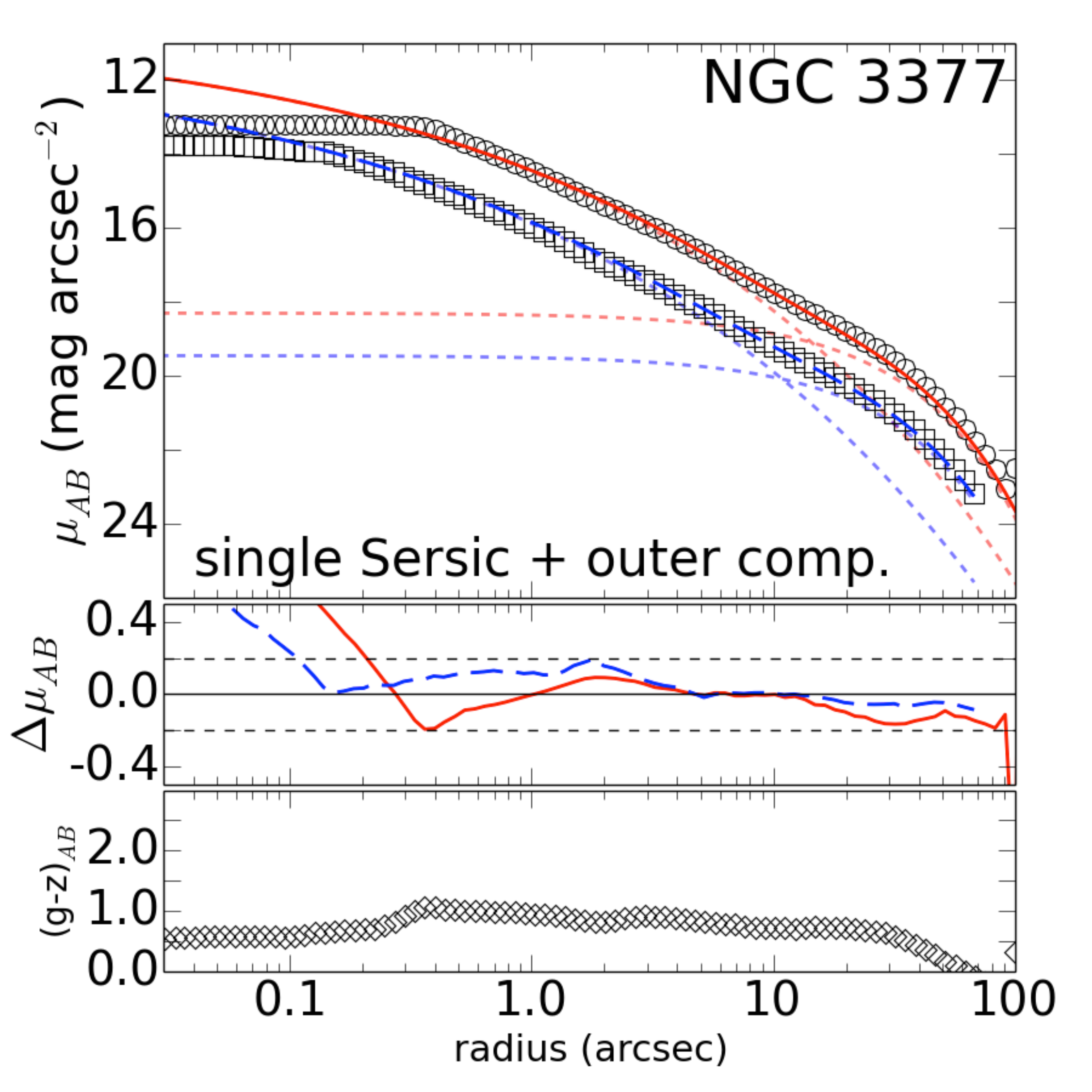} 

\end{tabular}
\caption{Light profiles for AMUSE-Field objects with HST coverage.  See Figure~\ref{lightpro1} for description.}
\label{lightpro3}
\end{figure*}

\begin{figure*}
\centering
\begin{tabular}{cc}

\includegraphics[scale=0.23]{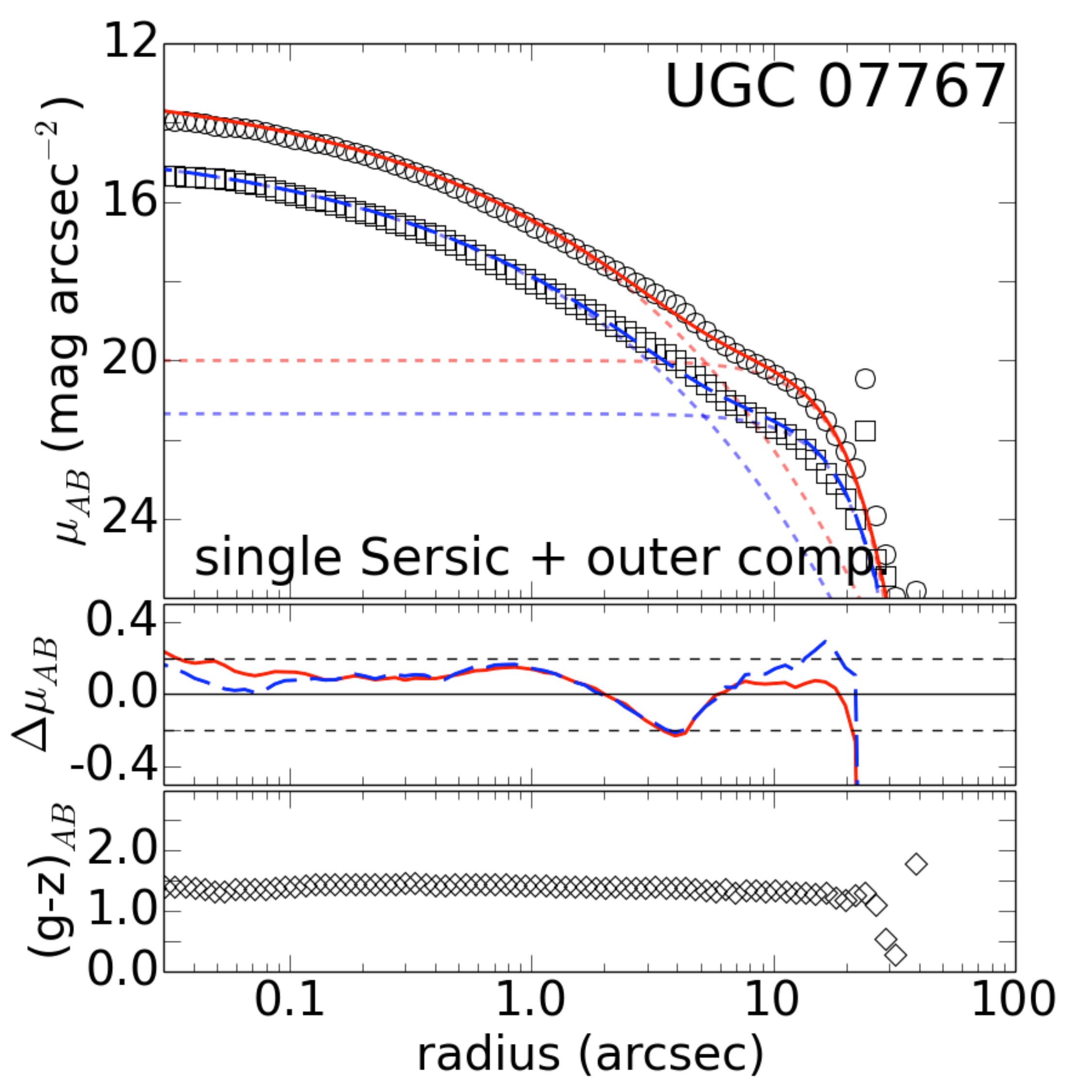} & \includegraphics[scale=0.23]{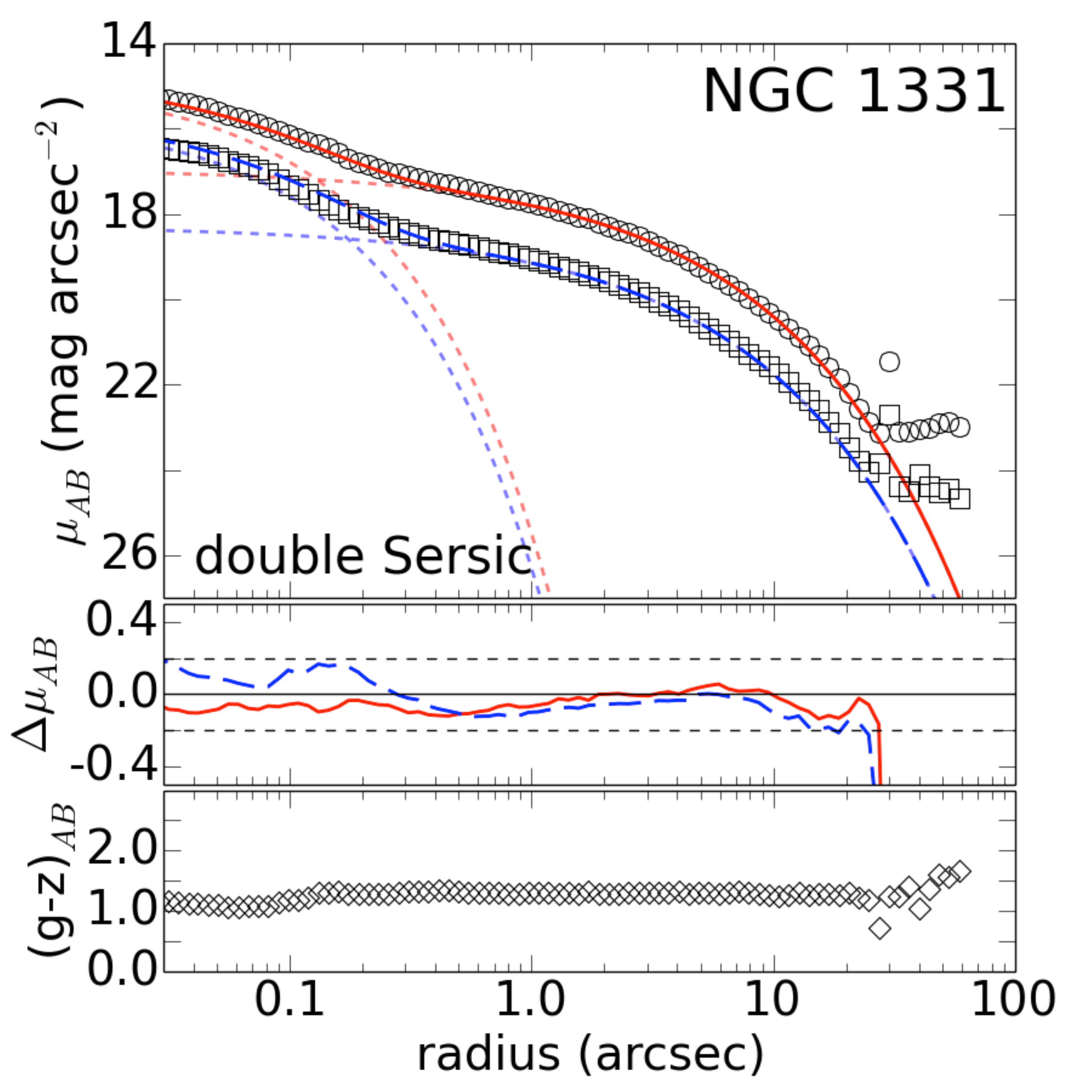} \\

\includegraphics[scale=0.23]{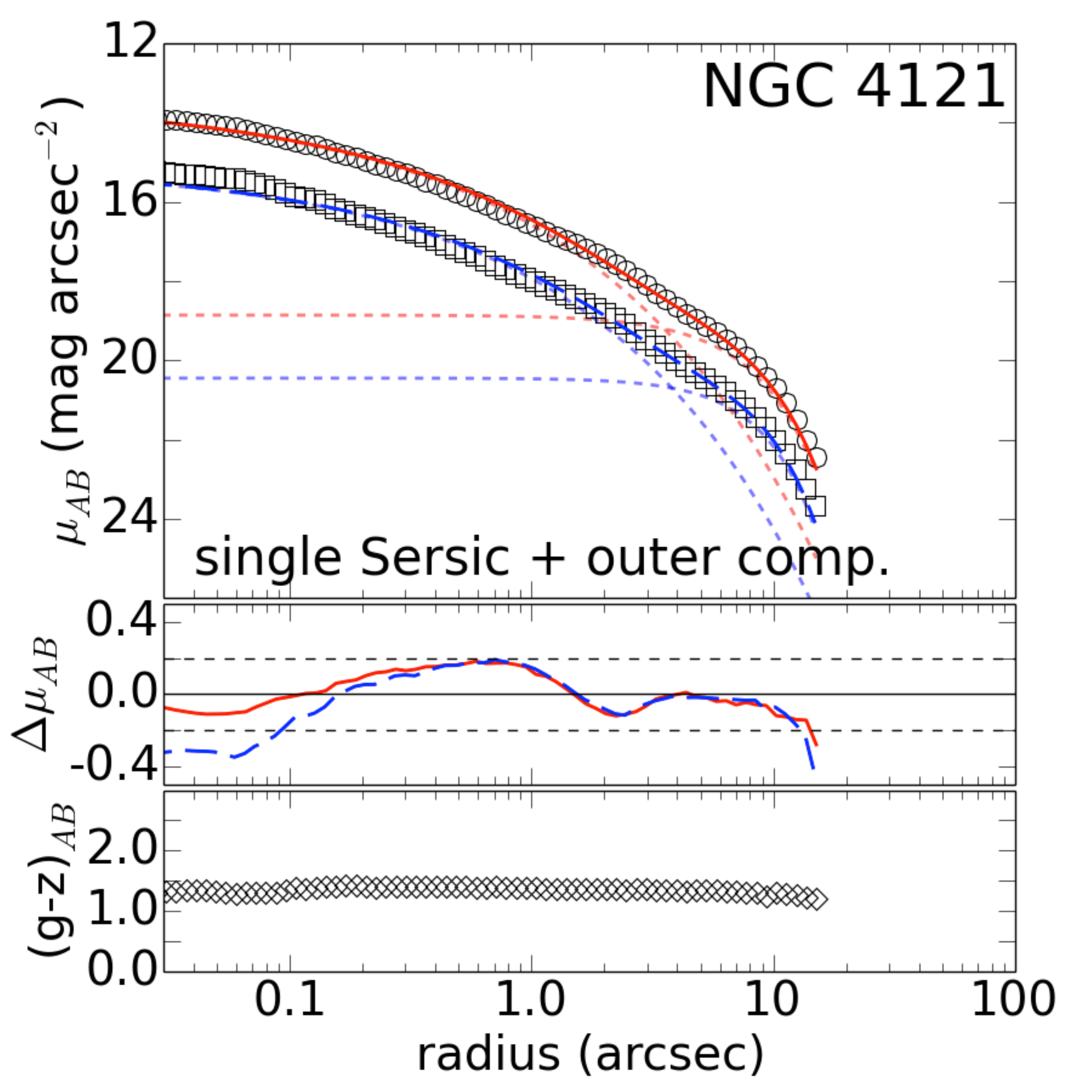} & \includegraphics[scale=0.23]{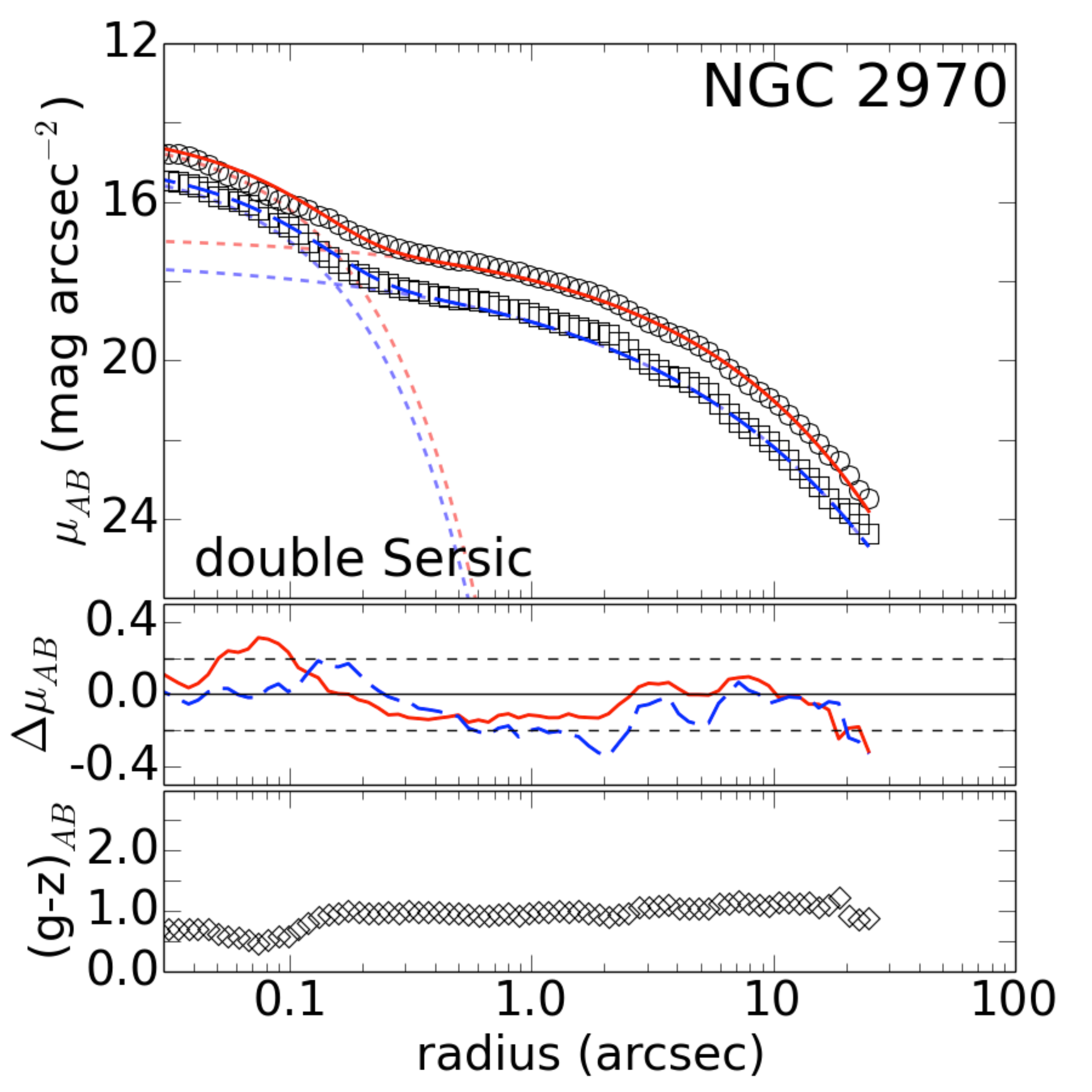} \\

\includegraphics[scale=0.23]{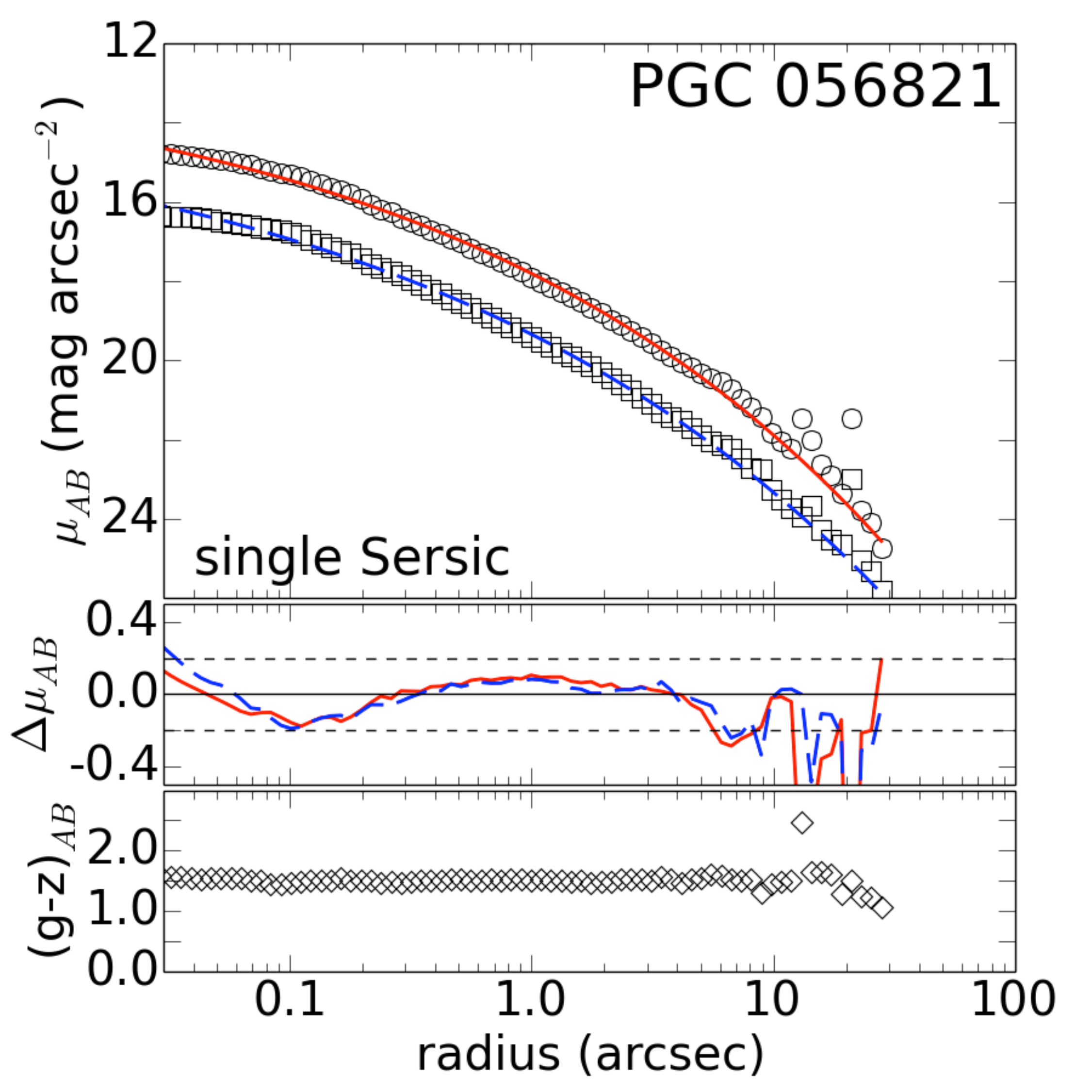} & 

\end{tabular}
\caption{Light profiles for AMUSE-Field objects with HST coverage.  See Figure~\ref{lightpro1} for description.}
\label{lightpro4}
\end{figure*}

Figures \ref{lightpro1} through \ref{lightpro4} show the 1-D surface brightness profiles, along with projected GALFIT models, best fit residuals, and color gradients for the 23 objects for which we were able to perform reliable photometry.

%%%%%%%%%%%%%%%

%%%

%%%

%%%

%%%%%%%%%%%%%%%

\section{Results}

\subsection{Nucleation Fraction}

Following \cite{2007ApJ...671.1456C}, we adopt the $\Delta_{0.02}$ parameter to quantify the degree of nucleation in our targets.  $\Delta_{0.02}$ is defined as $\log{L_{\rm g} / L_{\rm s}}$ where $L_{\rm g}$ is the total luminosity of the best-fit galaxy model inside a break radius $R_{\rm b}$, and $L_{\rm s}$ is the luminosity of just the outer S\'{e}rsic component in this region.  Break radius $R_{\rm b}$ is equal to $0.02R_{\rm e}$, which is where deviations from the outer S\'{e}rsic profile tend to occur.   Negative $\Delta_{0.02}$ values indicate nuclear light deficits; these ``cored" galaxies tend to be very luminous ellipticals \citep{Graham:2003ly}.  A positive $\Delta_{0.02}$ value indicates a nuclear light excess, i.e. a NSC.  

\begin{figure*}
\centering
\includegraphics[scale=0.45]{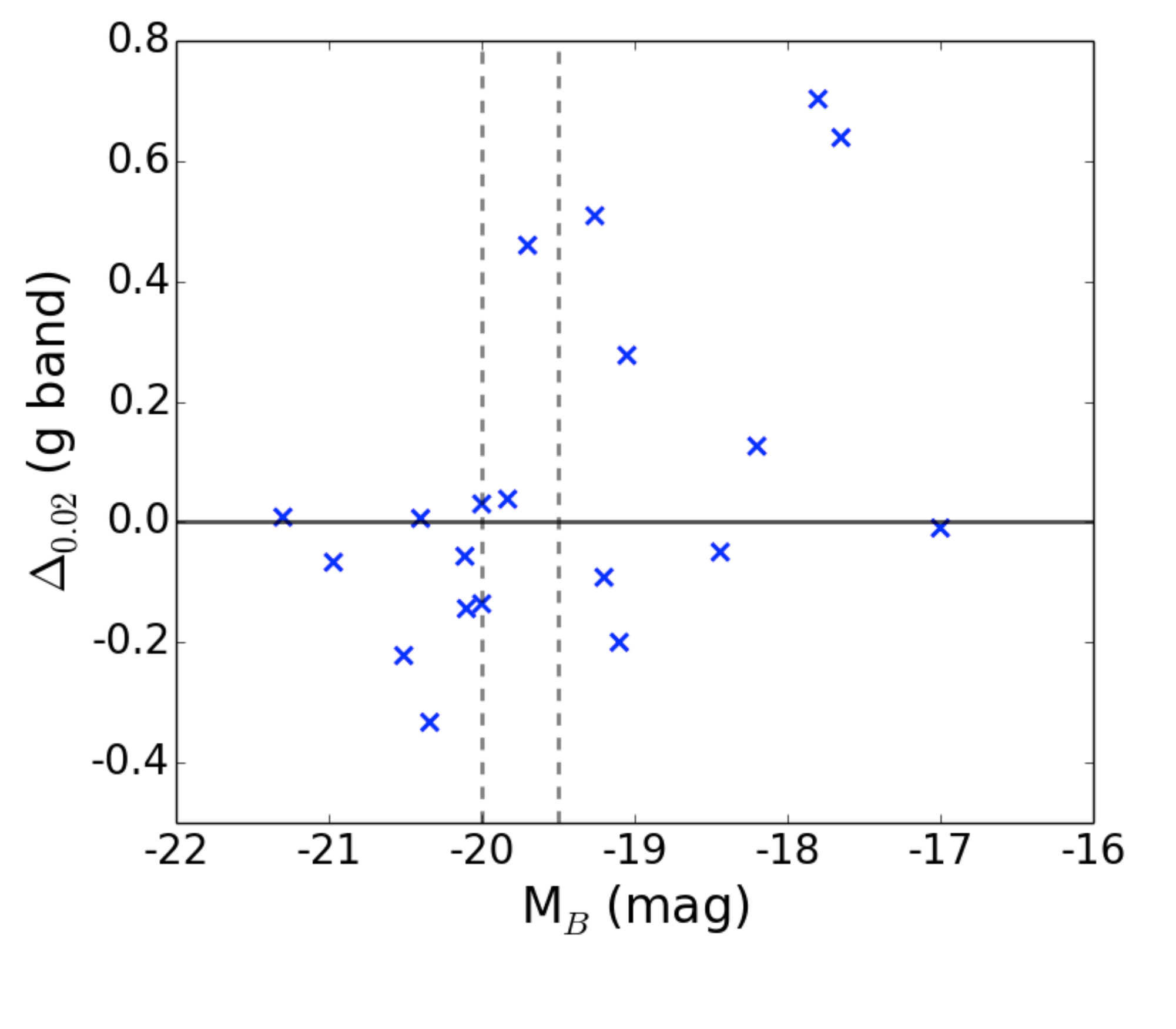} 
\caption{$\Delta_{0.02} = \log(L_{\rm g}/L_{\rm s})$ (as measured using g band data) versus absolute B-band magnitude.  Positive values of  $\Delta_{0.02}$ indicate there is a nuclear light excess, while negative values indicate a nuclear light deficit.  Dashed vertical lines bound the transition region from light deficit to excess identified by \cite{2007ApJ...671.1456C}.}
\label{delta02}
\end{figure*}

$\Delta_{0.02}$ is plotted against absolute B-band magnitude (from \citealt{:zr}) in Figure \ref{delta02} for 22 of the 23 objects in our analysis; we exclude NGC 4697 from this calculation because its nuclear disk precluded us from calculating an accurate  $\Delta_{0.02}$ value.  For objects with nuclear light deficits, we calculate $L_{\rm g}$ based on the flux inside $R_{b}$, as measured by ELLIPSE, as opposed to the best fit model, and use the outer S\'{e}rsic component to calculate $L_{\rm s}$. 
The Spearman rank for this relation, calculated using the $\Delta_{0.02}$ values measured from the g band data, is 0.36 with a p-value of 0.11, indicating a positive correlation between magnitude and $\Delta_{0.02}$ at the 89\% confidence level.  Unlike \cite{2007ApJ...671.1456C}, who find a consistent trend from nuclear light deficit to excess with a clearly defined transition region between $-20 < M_{\rm B} < -19.5$ mag, we do not observe such a sharp transition region for our sample. We check that the same qualitative results hold if the analysis is performed in the $z$ band. 

%%%%%%%%%%%%%
%%%%%%%%%%%%%

\begin{figure*}
\centering
\includegraphics[scale=0.60]{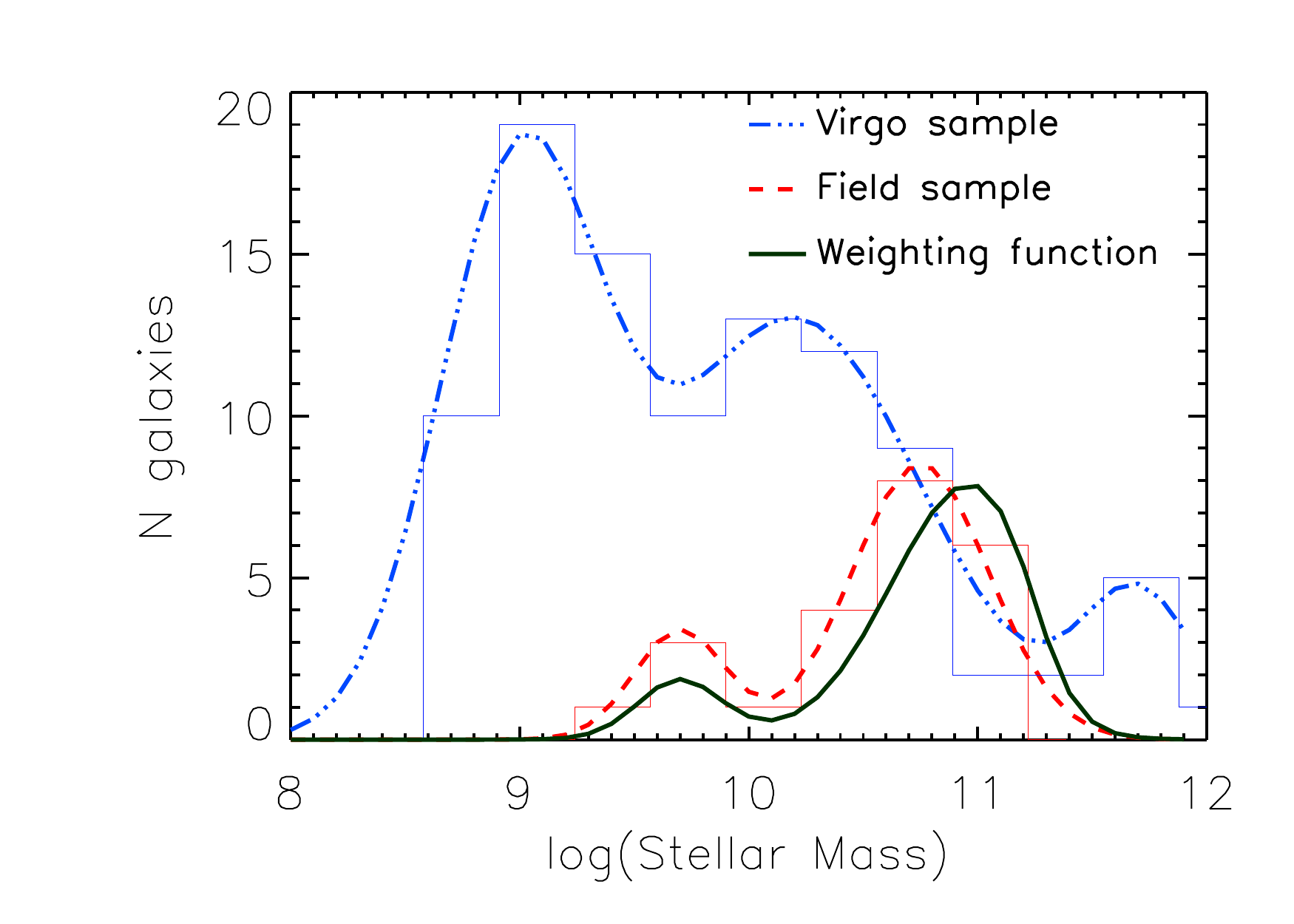}
\caption{Histograms of stellar mass distributions for the ACS VCS sample and Field sample with HST coverage.  We represent these histograms as the sum of Gaussians.  Also shown is the weighting function (arbitrary normalization) used to draw mass distribution matched subsamples from the Virgo Cluster; the weighting function is equivalent to the ratio of the Field to Virgo Gaussian representations.}
\label{massmatch}
\end{figure*}

Next, we compare our results for early-type field galaxies to those for the 100 early-type galaxies that compose the ACS VCS \citep{2004ApJS..153..223C}, in order to test whether the nucleation fraction has an environmental dependence.  In order to properly account for the different mass distributions across the two samples, we use the procedure outlined in Section 2.2 of \cite{:fg} to match the mass distributions, thus controlling for stellar mass (the mass distribution of the field sample is biased toward high stellar masses, since observations targeted objects with X-ray detections).  In brief, we represent the field and Virgo $M_{\star}$ distributions as a sum of Gaussian functions, then use a weighting function (equivalent to the ratio of the field to Virgo Gaussian representations) to draw subsamples from the Virgo sample that have the same mass distribution as the field sample.  Figure \ref{massmatch}  shows the two $M_{\star}$ distributions and their Gaussian representations, along with the weighting function that is used to draw subsamples from the Virgo cluster sample.  %
We draw 500 such subsamples of 23 Virgo galaxies each, and find that they contain, on average, 7.16 nucleated objects, corresponding to $30\%^{+17\%}_{-12\%}$ of galaxies (error given at 1$\sigma$ confidence level, \citealt{Gehrels:1986kx}).  For the field, we found six out of 23 objects to be nucleated, corresponding to $26\%^{+11\%}_{-18\%}$ (errors given at the 1$\sigma$ confidence level).  Poisson statistics shows that for an expected value of six nucleated galaxies, there is a 15\% chance of finding eight or more nucleated objects in a sample of 23.  This argues for no statistically significant difference in the nucleation fractions of the field and Virgo samples.

It is important to note that our fitting procedure differs from that of the ACS VCS; GALFIT performs 2-D modeling, while the ACS VCS fits directly to the 1-D surface brightness profile.  Additionally, the ACS VCS only fitted either a S\'{e}rsic or core-S\'{e}rsic profile to the galaxy surface brightness profile, while we sometimes include an extra component to the outer regions of the galaxy.  In order to explore whether this difference in method would result in differing classifications,  and if so, whether it would affect the soundness of our nucleation fraction comparison, we tested our fitting procedure on eight ACS VCS objects.  For seven out of eight objects, the classifications were consistent between methods.  One object, VCC 2095, which was classified by the ACS VCS as nucleated, was found not to be nucleated based on our GALFIT modeling.  However, we also failed to identify a NSC when fitting to the 1-D surface brightness profile, as done in \cite{2006ApJS..165...57C}.  We believe this particular discrepancy to be due to the ACS VCS supplementing the light profile fitting-based classifications with by-eye classifications.  To summarize, while there may be slight inconsistencies between classification methods, we expect the effect on the nucleation fraction to be small compared to the 1-$\sigma$ error bars, and not to affect the final conclusion that the nucleation fraction is consistent across environment.  

%%%%%%%%%%%%%
%%%%%%%%%%%%%

%
\subsection{Nucleation and X-ray Emission}

As discussed in Section 1, the X-ray luminosity threshold of the \cxo\ AMUSE surveys demands a careful assessment of the possible contamination of the nuclear X-ray signal from bright LMXBs, as opposed to low-luminosity SMBHs.  While \cite{:zr} {\it assumed} that the same level of nucleation (and thus contamination) measured in Virgo applied to the field galaxies, the ACS observations presented here enable us to verify this assumption by directly measuring the nucleation fraction.
While, in the absence of a NSC, the X-ray Luminosity Function (XLF) -- and thus expected number -- of LMXBs within the Chandra point spread function simply scales within the enclosed stellar mass (\citealt{:kj, :um}, and references therein), the presence of a NSC likely implies an enhanced LMXB contribution to the nuclear X-ray signal, as discussed below.  In order to quantify this effect, following \cite{:um} we adopt the functional shape of the XLF for LMXBs in globular clusters as estimated by \cite{:yf} \footnote{Number of LMXBs ($L_{\rm X} > 3.2 \times 10^{38} \rm{erg/s}$) $\propto$ $10^{0.90(g-z)}$ $r_{\rm h,cor}^{-2.22}$ ${M}^{1.24}$  , where $r_{\rm h,cor}$ is the half light radius in pc, and $M$ is the stellar mass in units of $10^{6}M_{\odot}$.}.  We expect this to be a conservative estimate of the probability of contamination for NSCs, as illustrated by the following discussion.

\cite{Pooley:2003ys} finds that the number of LMXBs in a dense stellar environments scales as the stellar encounter rate, $\Gamma$, to the 0.74 power; in turn, $\Gamma \propto \frac{\rho_{0}^{2} r_{c}^{3}}{\nu_{0}}$, where $\rho_{0}$ is the central density, $r_{\rm c}$ is the system core radius, and $\nu_{0}$ is the central velocity dispersion.  
While the NSCs have higher central densities compared to globulars (NSCs are about an order of magnitude more massive than the typical Milky Way globular cluster \citep{2005ApJ...618..237W}, the smaller core radii and the observed higher velocity dispersions for the NSCs suggest that globular clusters would have higher encounter rates, and thus a larger number of LMXBs (see \citealt{2005ApJ...618..237W} for velocity dispersions of NSCs; see \citealt{Harris:1996bh} for velocity dispersions of Milky Way globular clusters, see \citealt{2008ApJ...681.1136C}  for a comparison between the NSC and globular cluster profiles).  

Table~\ref{NSC_props} lists probabilities of contamination for objects that contain NSCs and have nuclear X-ray detections.  For reference, we point out that NGC 3384, which has the highest expected number of LMXBs of all the hybrid nuclei, is known to host a SMBH based on dynamical evidence  \citep{Gebhardt:2003gf}.  The listed probabilities have been incorporated in an accompanying paper (Miller \etal, submitted) where the Virgo and field samples combined are used to provide the first measurement of the SMBH occupation fraction in the local universe.

\section{Discussion and Conclusions}

Detailed morphological analysis of a sample of 100 early type galaxies in the Virgo cluster has shown evidence that the majority of galaxies with stellar mass below $10^{10}$ \msun\ host NSCs (e.g. Ferrarese \etal 2006a). 
In this paper, we explored the possibility of an environmental dependence of the nucleation fraction using dual-band (F475W \& F850LP) HST/ACS images of a sample of 28 field early type galaxies out of the 103 galaxies that compose the AMUSE-Field X-ray survey (Miller \etal 2012a).
After controlling for different stellar mass distributions, we found there to be no statistically significant difference between the nucleation fraction for field ($26\%^{+17\%}_{-11\%}$) and Virgo ($30\%^{+17\%}_{-12\%}$) early-type galaxies (from the ACS VCS survey).  Here, we discuss our results in the context of NSC formation theories, and compare the measured nucleation fractions with the active fractions as estimated from X-ray diagnostics. 

The mechanism behind the mass assembly and evolution of NSCs is still uncertain, but two prevalent formation theories have emerged.  The \textit{dissipationless} model suggests that NSCs form from the infall of globular clusters through dynamical friction \citep{1975ApJ...196..407T}, while the \textit{dissipative} model posits that NSCs form by gas accumulating at the center of a galaxy (via mergers, i.e. \citealt{Mihos:1994lq}, or transport of gas in a disk, i.e. \citealt{2004ApJ...605L..13M}) and forming stars \textit{in situ}.  Additionally, both processes may contribute to nucleus formation, and evidence for recurring episodes of star formation and/or distinct kinematics \citep{Rossa:2006pd, Seth:2008rt, Paudel:2011nx, Seth:2010ys} -- particularly in late-type galaxies -- implies that, even if globular cluster infall forms the bulk of the mass in the NSC, some gas needs to accrete to the center and form stars at later times.

Simulations show that in-falling globular clusters can form NSCs with global properties that generally match those of observed NSCs, and can reproduce scaling relations observed between NSCs and their host galaxies (\citealt{2008ApJ...681.1136C}, \citealt{Antonini:2012uq}, \citealt{2013ApJ...763...62A}, \citealt{Agarwal:2011fk}, \citealt{Gnedin:2013qy}).  This formation scenario is consistent with the observed radial distribution of the number of globular clusters near galaxy centers being flatter than that of the spheroidal stellar component with the same age and metallicity \citep{Capuzzo-Dolcetta:2009fj}, indicative of inner depletion of globular clusters via dynamical processes.
There is also evidence, particularly in late-type galaxies, for more recent gas accretion, from observations of multiple stellar populations (e.g. \citealt{2006ApJ...649..692W}, \citealt{Rossa:2006pd}) and kinematic features that cannot be reproduced solely by globular cluster infall \citep{2011MNRAS.418.2697H}.
Overall, while globular cluster infall is able to reproduce the bulk properties of NSCs (masses and radii), subsequent injection of gas seems to be necessary to explain the full spectrum of observed properties.        

Further information on the nature of nucleation comes from comparing different environments. For example, \cite{:fz} explored the nucleation fraction and NSC properties between the Virgo and Fornax early-type cluster members (through the ACS VCS and ACS Fornax Cluster Survey, FCS), and found them to be consistent with each other.  
They suggest that this agreement in nucleation fraction between the two different cluster environments -- with Fornax being substantially smaller, colder and denser than its Northern counterpart -- indicates that factors related to large scale environmental properties do not have large effects on the formation of NSCs in early-type galaxies.  
Although our sample size is smaller than those of both the VCS and FCS, our results confirm and strengthen this hypothesis, as we find no statistically significant difference in the nucleation fraction between the field and the clusters' samples. In addition, early-type galaxies in both cluster environments as well as the field demonstrate a trend in which their nuclei move from having nuclear light deficits to nuclear light excesses as galaxy luminosity decreases. 
 
\begin{figure*}
\centering
\includegraphics[scale=0.60]{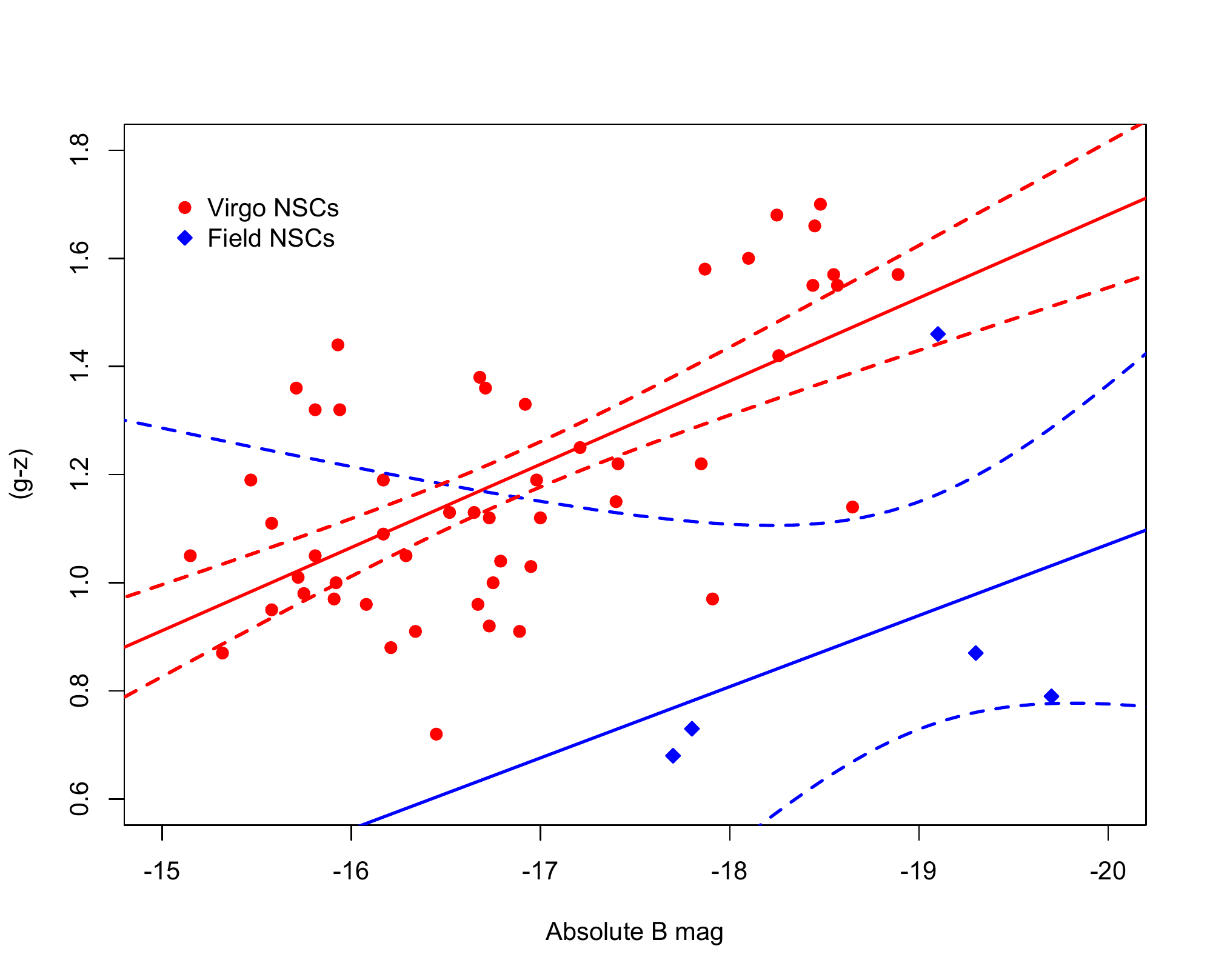}
\caption{(g-z) color of NSC versus host galaxy absolute B-band magnitude.  Red circles correspond to Virgo Cluster galaxies; blue diamonds to Field galaxies. Shown are the results of a linear regression analysis to test the presence of a relation of the form ($g-z$)$-1.2=a+b($ $M_{b}$). A statistically significant correlation is found for the Virgo NSCs (with best-fit slope $\beta=-0.15\pm0.03$, at 1$\sigma$), while no significant correlation is found for the field. 
}
\label{Mb_gz}
\end{figure*}

\begin{figure*}
\centering
\includegraphics[scale=0.40]{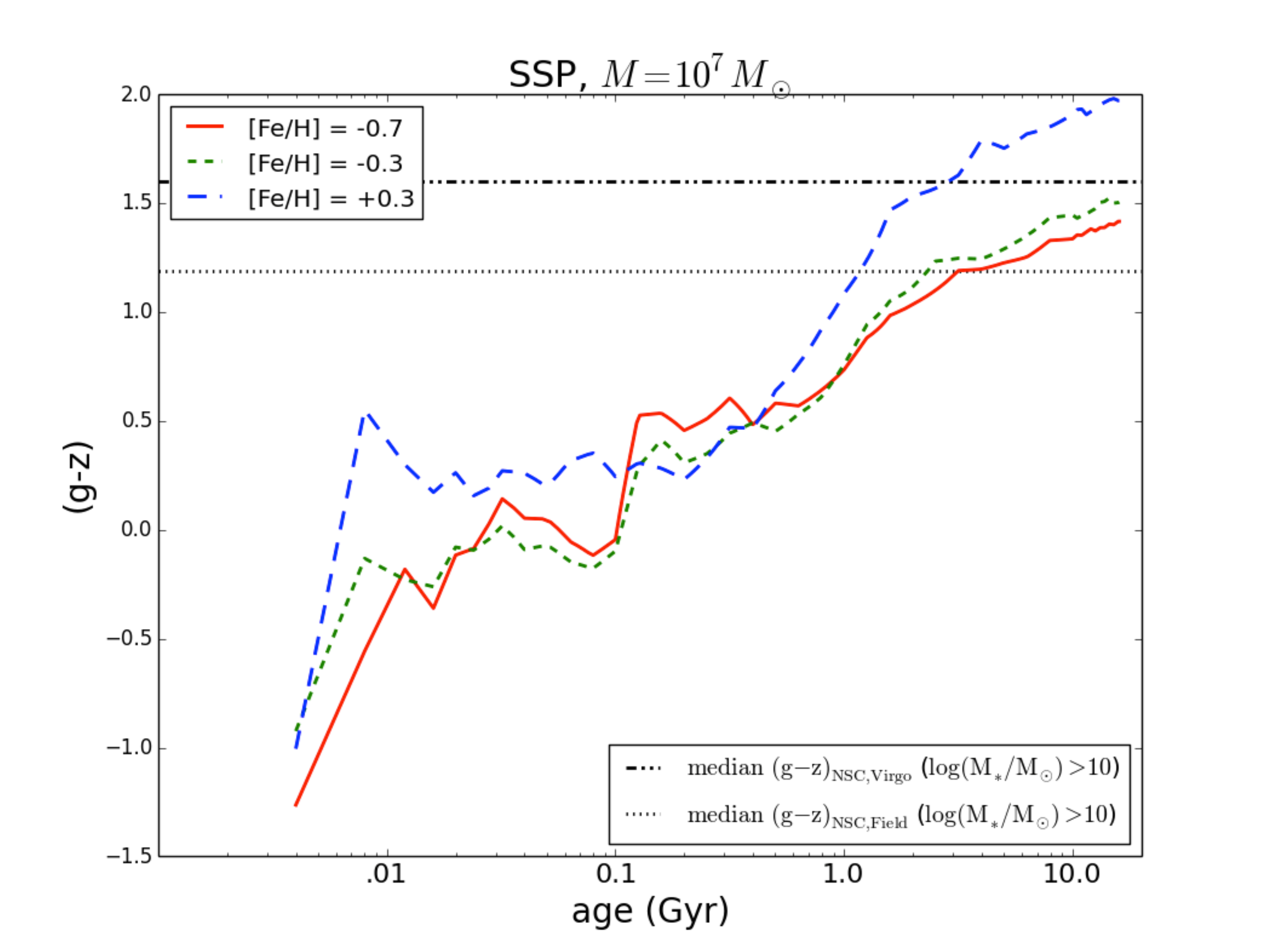}
\caption{A simple model for the evolution of the (g-z) color of a single stellar population with stellar mass \mstar = $10^{7}$ \msun.  Plotted for comparison are the median colors of the Virgo and Field NSCs in galaxies with $\log$(\mstar/\msun) $ > 10.0$.  }
\label{starpop}
\end{figure*}

In Figure~\ref{Mb_gz}, we explore whether environmental differences are reflected in the colors of field and Virgo NSCs.  
As discussed by \cite{:fz}, there is a tendency for Virgo NSCs (displayed as red filled circles) to become redder with increasing host galaxy mass/luminosity (see the caption of Figure~\ref{Mb_gz} for a quantitative analysis).  According to single stellar population models  \citep{Kotulla:2009ul}, the extremely red colors of the more massive Virgo NSCs (with (\textit{g-z}) as high as 1.7) can only be achieved in less than a Hubble time for super-solar metallicities (see Figure~\ref{starpop}). Thus, \cite{:fz} interpret these results as being suggestive of two main channels of growth for NSCs, with low mass nuclei being primarily assembled via globular cluster infall, and higher mass nuclei growing further through subsequent accretion, mergers, and/or tidal torques involving metal-enriched gas.
Our sample size is too small to ascertain whether field NSCs (displayed as blue triangles in Figure~\ref{Mb_gz}) show the same reddening with host stellar mass.  If the higher mass Virgo nuclei do redden as a result of growth through the aforementioned mechanisms, we might expect that NSCs in the field (where events like mergers and tidal torques are less common) show a more constant (\textit{g-z}) color as a function of stellar mass. 
%

%%%%%%%%%%% COLOR VS. MAG %%%%%%%%%%%

%%%%%%%%%%% STELLAR POP EV %%%%%%%%%%%

%%%%%%%%%%%%%

At the same time, highly sub-Eddington SMBH activity, as indicated by X-ray observations, changes with environment.  The higher incidence of nuclear X-ray activity in the field sample (50\% $\pm$ 7\% versus 32\% $\pm$ 6\%), and the tendency of field galaxies towards marginally higher X-ray luminosity (by 0.38 $\pm$ 0.14 dex) for a given host stellar mass, has been interpreted by Miller \etal (2012b) as due to the field galaxies, and their nuclear SMBHs, having access to greater cold gas reservoirs than the cluster galaxies, possibly as a result of enhanced ram pressure in a cluster environment.  

Combined, the X-ray and optical diagnostics point toward the following scenario:  In the field NSCs, small amounts of gas, possibly from residual star formation, has the potential to keep feeding a nuclear SMBH well after the major NSC mass assembly has occurred.  In contrast, gas is more likely to be depleted from cluster members due to ram pressure stripping.  Its important to note that the colors of the reddest Virgo NSCs still imply a stellar population that is older than $\sim 1$ Gyr (Figure~\ref{starpop}).  This is comparable to the crossing time for a massive galaxy cluster, making it likely that gas could have been stripped between the last episode of star formation and the present day, decreasing the amount of gas available to funnel to the nucleus.  
 
Complementary questions about the interplay between SMBHs, NSCs and globular clusters can potentially be addressed by exploring their relative spatial distribution within clusters. E.g., \cite{2008ApJ...681..197P} presents evidence for an environmental dependence of the specific frequency\footnote{Defined as the number of globular clusters normalized to a galaxy luminosity of $M_{\rm V} = -15$ \citep{Harris:1981ul}}  of globular clusters.  They find that Virgo dwarf galaxies with high specific frequencies tend to reside within 1 Mpc of  M87,  implying that globular cluster formation may be biased toward denser environments.  Although \cite{Binggeli:1987qf} found that nucleated galaxies in Virgo are more centrally concentrated than non-nucleated galaxies, \cite{2006ApJS..165...57C} did not replicate this trend, and attribute this finding to the surface brightness limit of \cite{Binggeli:1987qf}.  However, \cite{2006ApJS..165...57C} used only galaxies with $B_{\rm T} \ge 13.7$ to test this result, corresponding to 40 nucleated galaxies and just 4 non-nucleated galaxies.  Whether nucleated galaxies are more centrally concentrated than non-nucleated galaxies, whether this aligns with the concentration of galaxies with higher frequencies of globular clusters, and whether these correlate at all with nuclear SMBH accretion, warrants further investigation.   \\

V.F.B. is supported by the National Science Foundation Graduate Research Fellowship Program grant DGE 1256260.  This work was supported in part by the National Science Foundation under grant no. NSF PHY11-25915 and by NASA through grant HST-GO-12591 from the Space Telescope Science Institute, which is operated by the Association of Universities for Research in Astronomy, Inc., under NASA contract NAS 5-26555.  J.H.W. acknowledges support by the National Research Foundation of Korea (NRF) grant funded by the Korea government (MEST; No. 2012-006087).  V.F.B. thanks Oleg Gnedin for useful suggestions. \clearpage

\begin{table*}
\caption{Galaxy Properties}
\label{OptProps}
\centering
\begin{tabular}{l c c c c c c c }
\hline
\hline
Object Name & R.A. & Decl.  & Distance & $M_{\rm g}$ & $M_{\ast}$ & $(g-z)_{\rm AB}$  & Nucleated?\\
  &  (deg) & (deg)  & (Mpc) & (mags) & ($M_{\odot}$) & (mags) &  \\
\hline
NGC 4125 & 182.025000 & 65.174167 & 23.7 & -21.17 & 11.2 & 1.37 & n   \\
NGC 3585  & 168.321250 & -26.754722 & 19.9 & -20.94 & 11.1 & 1.32 & n   \\
NGC 4036 & 180.361667 & 61.895833 & 24.2 & -20.7 & 11.0 & 1.41 & n    \\
NGC 4291 & 185.075833 & 75.370833 & 32.2 & -20.67 & 10.9 & 1.23 & n  \\
NGC 1340 & 52.082083 & -31.068056 & 20.6 & -20.45 & 10.9 & 1.32 & y   \\
NGC 4278 & 185.028333 & 29.280833 & 18.5 & -20.40 & 10.8 & 1.39 & n   \\
NGC 5831 & 226.029167 & 1.220000 & 26.9 & -20.25 & 10.6 & 1.23 & n   \\
NGC 4697  & 192.149583 & -5.800833 & 12.2 & -20.11 & 10.6 & 1.37 & n   \\
NGC 3115 & 151.308333 & -7.718611 & 9.7 & -20.07 & 10.7 & 1.47 & n  \\
NGC 5582 & 215.179583 & 39.693611 & 28.2 & -20.02 & 10.6 & 1.14 & n   \\
NGC 3379 & 161.956667 & 12.581667 & 11.1 & -19.83 & 10.7 & 1.46 & n   \\
NGC 1439 & 56.208333 & -21.920556 & 26.4 & -19.83 & 10.6 & 1.35 & n   \\
NGC 5845 & 226.503333 & 1.633889 & 32.7 & -19.82 & 10.7 & 1.42 & n   \\
NGC 1426 & 55.704583 & -22.108333 & 23.3 & -19.66 & 10.6 & 1.37 & y    \\
NGC 4648 & 190.435000 & 74.420833 & 25.4$^{\rm a}$ & -19.63 & 10.4 & 1.27 & n   \\
NGC 3384 & 162.070417 & 12.629167 & 9.2 & -19.24 & 10.4 & 1.37 & y    \\
NGC 1172 & 45.400000 & -14.836667 & 22.0 & -19.09 & 10.3 & 1.30 & y   \\
NGC 3377 & 161.926250 & 13.985833 & 10.2 & -18.94 & 10.3 & 1.31 & n  \\
UGC 07767 & 188.885000 & 73.674722 & 27.5 & -18.57 & 10.0 & 1.21 & n  \\
NGC 1331 & 51.617917 & -21.355278 & 22.9 & -18.14 & 9.8 & 1.06 & y    \\
NGC 4121 & 181.985833 & 65.113889 & 24.8$^{\rm a}$ & -18.08 & 9.8 & 1.11 & n   \\
NGC 2970 & 145.879583 & 31.976944 & 25.9$^{\rm a}$ & -17.86 & 9.6 & 0.96 & y  \\
PGC 056821 & 240.697917 & 19.787222 & 27.0$^{\rm a}$ & -17.18 & 9.5 & 1.23 & n   \\
NGC 3265 & 157.778333 & 28.796667 & 23.0$^{\rm a}$  & -- & -- & --  & --   \\
NGC 3073 & 150.217083 & 55.618889 & 33.4 & -- & -- & -- & -- \\
NGC 1370 & 53.810833 & -20.373611 & 13.2$^{\rm a}$  & -- & -- & -- & --   \\
NGC 0855 & 33.514583 & 27.877222 & 12.96 & -- & -- & -- & --  \\
ESO 540-014 & 10.298750 & -21.131667 & 22.4$^{\rm a}$  & -- & -- & --  & --  \\

\hline

\end{tabular}

\textbf{Notes. }R.A. and Decl. are taken directly from the HyperLeda database.  Distances are calculated from the redshift-independent mod0 distance modulus in HyperLeda ($^{\rm a}$For objects lacking a mod0 distance modulus, the modz redshift-based distance modulus was used).  $M_{\rm g}$, $M_{\star}$, and color are determined as described in Section 2.
\end{table*}

\begin{table*}
\caption{Nuclear Star Cluster Properties}
\label{NSC_props}
\centering
\begin{tabular}{c c c c c c c }
Host Galaxy & $r_{\rm h}$ & $\log(M_{\rm NSC}$) & $(g-z)$&  $M_{\rm NSC}/M_{\rm gal}$ & N($>L_{\rm X}$) & 1-$P_{\rm X}$ \\
& (pc) & ($M_{\odot}$) & (mag) & & & (\%) \\
(1) & (2) & (3) & (4) & (5) & (6) & (7) \\
\hline
NGC 3384 & 8.0 & 7.8 & 1.46 & .003 & 0.37 & 30.9 \\
NGC 1340 & 131.8 & 8.9 & 1.32 & .010 & 0.10 & 9.5 \\
NGC 1426 & 57.6 & 8.6 & 0.81 & .010 & 0.42 & 34.3 \\
NGC 1172 & 26.1 & 8.3 & 1.05 & .010 & 0.16 & 14.8 \\
NGC 2970 & 11.3 & 7.5 & 0.67 & .008 & 0.01 &  1.0 \\
NGC 1331 & 16.2 & 7.3 & 0.69 & .003 & 0.02 & 1.9 \\
\hline
\end{tabular}

\textbf{Notes.} Column 2: Half light radius.  Column 3: Mass of NSC.  Column 4: (g-z) color of NSC. Column 5: Fraction of host galaxy stellar mass contained in the NSC. Column 6: For objects with nuclear X-ray detections, number of expected LMXBs with $\log(L_{\rm X}) > 38.2$ within the \cxo~PSF.  Column 7: For galaxies with nuclear X-ray detections, probability that the Chandra PSF is contaminated by an LMXB of $\log(L_{\rm X}) > 38.2$.  
\end{table*}

\appendix

\section{Additional Information on Individual Objects}

\subsection*{ESO 540-014}

This object was misclassified as an early-type galaxy based on observations from the ground.  Our observations reveal clumps of star formation and an irregular morphology with no clear nucleus.  This object may be considered a dE/dIrr object, as defined in Ferrarese et al. 2006.  This object was excluded from analysis.

\subsection*{NGC 0855}

This object is highly irregular, with many clumps of star formation extending in a bar across the galaxy.  There is also a great deal of dust contamination. This object may be considered a dE/dIrr object, as defined in Ferrarese et al. 2006.  This object was excluded from analysis.

\subsection*{NGC 1172}

NGC 1172 is elliptical, with dust lanes stretching across the galaxy and diffuse dust throughout.  NGC 1172 is nucleated and is well fit by a double S\'{e}rsic profile with the addition of low ($n\sim0.5$) S\'{e}rsic index component to fit the outer regions. 

\subsection*{NGC 1331}

NGC 1331 is an elliptical galaxy with a nuclear star cluster.  This object's light profile is well fit by a double-S\'{e}rsic profile.  The nucleus is slightly bluer than the outer regions of the galaxy.  In fitting with GALFIT, the S\'{e}rsic index of the nucleus had to be held fixed.  

\subsection*{NGC 1340}

NGC 1340 is nucleated and is well fit by a double S\'{e}rsic profile.  

\subsection*{NGC 1370}

NGC 1370 has an extremely dusty torus.  Dust dominates a significant portion of the galaxy in g band, and is visible around the nucleus in z band.  This object was excluded from analysis.

\subsection*{NGC 1426}

NGC 1426 is nucleated and is best fit by a double-S\'{e}rsic profile.

\subsection*{NGC 1439}

NGC1439 has a dusty disk around its nucleus extending about .5".  The majority of the galaxy is well fit by a single S\'{e}rsic profile, but an extra component is necessary to fit the region past ~20". 

\subsection*{NGC 2970}

NGC 2970 has a faint spiral structure, perhaps from a merger.  It is the bluest object in our sample with a $(g-z)$ color of 0.96.  NGC 2970 has a nuclear star cluster and is well fit by a double-S\'{e}rsic profile.

\subsection*{NGC 3073}

NGC 3073 suffers from dust contamination throughout the galaxy.  It is particularly problematic in g band, but the dust is also visible in z band.  This object was excluded from analysis.

\subsection*{NGC 3115}

NGC 3115 is a highly edge-on S0 galaxy.  The light profile was fit by a S\'{e}rsic profile plus an exponential disk.

\subsection*{NGC 3265}

NGC 3265 appears to have very diffuse spiral arms and is highly contaminated by dust.  This object was excluded from analysis.

\subsection*{NGC 3585}

NGC 3585 appears to have a disk. The light profile was fit by a S\'{e}rsic profile plus an outer exponential disk component.

\subsection*{NGC 3377}

NGC 3377 has an extremely depleted core, which is also very blue.  Excluding the central region, we can fit the light profile of this galaxy with a S\'{e}rsic profile plus an outer disk component.  NGC 3377 has a dust lane as well as diffuse dust contamination in the g band.  There is a steep color gradient from the nucleus to the outer regions of the galaxy, with the $g-z$ color getting redder with increasing radius.

\subsection*{NGC 3379}

NGC 3379 had a chip gap across the nucleus, but we were able to correct for it.  There appears to be a disk in the central 
2" but the chip gap covers part of it.  This galaxy is well fit by a S\'{e}rsic profile plus an outer disk-like component.

\subsection*{NGC 3384}

NGC 3384 required a multiple component fit for the outer galaxy, and was found to have a nuclear star cluster.  This galaxy has been referenced in the literature as having both a NSC and SMBH \citep{2009MNRAS.397.2148G}.

\subsection*{NGC 4036}

NGC 4036 suffers from significant dust contamination, including some contamination surrounding the nucleus.  The light profile was fit by a S\'{e}rsic component plus an outer exponential disk component.

\subsection*{NGC 4121}

NGC 4121 is well fit by a S\'{e}rsic component for most of the galaxy plus an outer exponential disk component.

\subsection*{NGC 4125}

NGC 4125 suffers from diffuse dust contamination across most of the galaxy and across the nucleus.  The light profile is well fit by a single S\'{e}rsic component.

\subsection*{NGC 4278}

NGC 4278 is well fit by a single S\'{e}rsic component.

\subsection*{NGC 4291}

NGC 4291 has a depleted core, but the outer regions of the galaxy are well fit by a single S\'{e}rsic profile.

\subsection*{NGC 4648}

The light profile of NGC 4648 is well fit by a S\'{e}rsic component for most of the galaxy plus an outer exponential disk component.

\subsection*{NGC 4697}

NGC 4697 has a disk in the center which complicates light profile fitting.  The disk extends from 0.4" to 4.0", corresponding to a dip in the light profile visible in this region.  While NGC 4697's light profile can be fit with a double S\'{e}rsic profile, we are not convinced it is truly nucleated as opposed to appearing to have a central light excess with respect to the nuclear disk.

\subsection*{NGC 5582}

The light profile of NGC 5582 is well fit by a single S\'{e}rsic profile.

\subsection*{NGC 5831}

The light profile of NGC 5831 is well fit by a single S\'{e}rsic profile.

\subsection*{NGC 5845}

NGC 5845 has a dusty disk in the center.  Its light profile is well fit by a single S\'{e}rsic profile.

\subsection*{PGC 056821}

The light profile of PGC 056821 is well fit by a single S\'{e}rsic profile.  The images of PGC 056821 taken in the F850LP filter suffer from streaks of scattered light from a nearby star.  Some of these streaks cut across the galaxy.  These were corrected for by masking them, creating a model using the IRAF ELLIPSE and BMODEL tasks, and filling in those regions with the model.

\subsection*{UGC 07767}

The light profile of UGC 07767 is well fit by a S\'{e}rsic component plus an outer exponential disk component.

\end{document}